\documentclass[twocolumn,nodate,eqsecnum,showpacs,preprintnumbers,nofootinbib,amsmath,amssymb,aps,prd]{revtex4-1}

\usepackage{graphicx}
\usepackage{amsmath,amssymb}

\usepackage[english]{babel}
\usepackage{graphics}
\usepackage{color}
\usepackage{wrapfig}
\usepackage{epsfig}
\usepackage{pst-grad} 
\usepackage{times}

\def\b{$\bullet\,\,\,$}
\def\f{\frac}
\def\be{\begin{equation}}
\def\ee{\end{equation}}
\def\ba{\begin{eqnarray}}
\def\ea{\end{eqnarray}}

\def\scri{\mathcal{I}}

\def\szpr{\mathcal{I}^{o+}_{\rm R}}
\def\szpl{\mathcal{I}^{o+}_{\rm L}}
\def\szmr{\mathcal{I}^{o-}_{\rm R}}
\def\szml{\mathcal{I}^{o-}_{\rm L}}
\def\s1pr{\mathcal{I}^{1+}_{\rm R}}

\def\spr{\mathcal{I}^{+}_{\rm R}}
\def\spl{\mathcal{I}^{+}_{\rm L}}
\def\sml{\mathcal{I}^{-}_{\rm L}}
\def\smr{\mathcal{I}^{-}_{\rm R}}
\def\sprb{\bar{\mathcal{I}}^{+}_{\rm R}}
\def\mbt{M_{\rm Bondi}^{\rm Trad}}
\def\mbatv{M_{\rm Bondi}^{\rm ATV}}
\def\ADM{{\rm ADM}}

\def\Bondi{{\rm Bondi}}
\def\fatv{F^{\rm ATV}}
\def\ftrad{F^{\rm Trad}}
\def\GDH{{\rm GDH}}
\def\N{\bar{N}}

\def\h{\hat}
\def\t{\tilde}
\def\ub{\underbar}
\def\ul{\underline}
\def\b{\bar}

\def\Dp{\partial_{+}}
\def\Dm{\partial_{-}}
\def\dd{{\rm d}}

\def\a{\mathbf{a}}

\def\k{\kappa}
\def\R{\mathbb{R}}

\begin{document}

\title{Evaporation of 2-Dimensional Black Holes}

\author{Abhay Ashtekar${}^{1}$ }
\email{ashtekar@gravity.psu.edu}
\author{Frans Pretorius ${}^{2}$}
\email{fpretori@princeton.edu}
\author{Fethi M. Ramazano\u{g}lu${}^{2}$}
    \email{framazan@princeton.edu}
\affiliation{${}^1$\! Institute for Gravitation and the Cosmos \&
     Physics Department, Penn State, University Park, PA 16802,
    USA\\
${}^2$\! Department of Physics, Princeton University, 08544,
Princeton, NJ, USA }	

\begin{abstract}

We present a detailed analysis of results from a new study of the
quantum evaporation of Callan-Giddings-Harvey-Strominger (CGHS)
black holes within the mean-field approximation. This semi-classical
theory incorporates back reaction. Our analytical and numerical
calculations show that, while some of the assumptions underlying the
standard evaporation paradigm are borne out, several are not. One of
the anticipated properties we confirm is that the semi-classical
space-time is asymptotically flat at right future null infinity,
$\spr$, yet incomplete in the sense that null observers reach a
future Cauchy horizon in finite affine time. Unexpected behavior
includes that the Bondi mass traditionally used in the literature
can become negative even when the area of the horizon is
macroscopic; an improved Bondi mass remains positive until the end
of semi-classical evaporation, yet the final value can be
arbitrarily large relative to the Planck mass; and the flux of the
quantum radiation at $\spr$ is non-thermal even when the horizon
area is large compared to the Planck scale. Furthermore, if the
black hole is initially macroscopic, the evaporation process
exhibits remarkable universal properties. Although the literature on
CGHS black holes is quite rich, these features had escaped previous
analyses, in part because of lack of required numerical precision,
and in part due to misinterpretation of certain properties and
symmetries of the model. Finally, our results provide support for
the full quantum scenario recently developed by Ashtekar, Taveras
and Varadarajan, and also offer a number of interesting problems to
the mathematical relativity and geometric analysis communities.

\end{abstract}

\maketitle

\section{Introduction}
\label{s1}

Although literature on the quantum nature of black holes is vast,
many important questions on the dynamics of their evaporation still
remain unanswered. This is true even for 2-dimensional models,
introduced some twenty years  ago~\cite{cghs,reviews,old}. In this paper,
as a follow up to ~\cite{apr}, we present a detailed analysis of the
semi-classical dynamics of 2-dimensional,
Callan-Giddings-Harvey-Strominger (CGHS) black holes \cite{cghs},
using a combination of analytical and high precision numerical
methods. This model has been studied extensively in the
past~\cite{reviews}. Yet we find new and rather remarkable behavior
for classes of black holes where the collapse is prompt and the
Arnowitt-Deser-Misner (ADM) mass is large: As these black holes
evaporate, after an initial transient phase, dynamics of various
physically interesting quantities flow to universal curves. In
addition, this analysis brings out certain unforeseen limitations of
the standard paradigm that has been used to discuss black hole
evaporation for some two decades. The overall situation bares
interesting parallels to the discovery of critical phenomena at the
threshold of gravitational collapse in classical general relativity
\cite{mc,gmg}; at the time it was assumed that the spherically
symmetric gravitational collapse problem was essentially ``solved'',
yet all earlier work had missed this very rich and profound property
of gravitational collapse. Also, it served as a new example of
rather simple, universal behavior that dynamically emerges from a
complicated system of non-linear partial differential equations.
Here, though the physical scenario is different, a similar
unexpected universal behavior arises. This universality strongly
suggests that, although the S-matrix is very likely to be unitary,
all the information in the infalling matter will not be imprinted in
the outgoing quantum state. As discussed in detail in the paper,
this mismatch between unitarity and information recovery is a
peculiarity of 2 dimensions.

Our investigation is carried out within the mean-field approximation
of \cite{atv} in which the black hole formation and evaporation is
described entirely in terms of non-linear, partial differential
equations (PDEs). Thus there will be no Hilbert spaces,
operators or path integrals in this paper.%
\footnote{This paper complements a companion article \cite{vat}
which is primarily devoted to quantum issues, particularly that of
potential information loss, and another companion article \cite{rp}
devoted to the details of numerical simulations.}
The focus is rather on the consequences of the geometrical PDEs
governing black hole evaporation in the semi-classical regime, and
the intended audience is mathematical and numerical relativists.
Therefore we will start with an introductory review of the CGHS
model---the classical field theory in Sec.~\ref{class_intro} and
semi-classical corrections in Sec.~\ref{semi_intro}---and then
summarize the results and outline the rest of the paper in
Sec.~\ref{summary_intro}.

\subsection{The Classical CGHS Model}\label{class_intro}

Consider first the spherical collapse of a massless scalar
field $f$ in 4 space-time dimensions. Mathematically, it is
convenient to write the coordinate $r$ which measures the
physical radius of metric 2-spheres as $r = e^{-\phi}/ \k$
where $\k$ is a constant with dimensions of inverse length. The
space-time metric\, ${}^4g_{ab}$ can then be expressed as
\be {}^4g_{ab} = \ub{g}_{ab} + r^2 s_{ab} := \ub{g}_{ab} +
\f{e^{-2\phi}}{\kappa^2}\,s_{ab}\,  , \ee
where $s_{ab}$ is the unit 2-sphere metric and $\ub{g}_{ab}$ is the
2-metric in the r-t plane. In terms of these fields, the action for
this Einstein-Klein-Gordon sector can be written as
\ba \label{action1} \tilde{S}(\ub{g},\phi, f)\!\! &=&\!\! \f{1}
{8\pi\,G_4}\, \f{4 \pi}{\k^2}  {\textstyle{\int}} d^2x
\sqrt{|\ub{g}|}\, e^{-2\phi} \,(\ub{R} + {\bf 2} \nabla^a\phi
\nabla_a \phi \nonumber\\
&+& {\bf 2 e^{-2\phi}} \kappa^2) - {\textstyle \f{1}{2}\,{\int}}
d^2x \sqrt{|\ub{g}|}\, {\bf e^{-\phi}} \nabla^a f \nabla_a f \ea
where $G_4$ is the 4-dimensional Newton's constant, $\nabla$ is
the derivative operator and $\ub{R}$ the scalar curvature of
the 2-metric $\ub{g}_{ab}$. The CGHS model, by contrast refers
to gravitational collapse of a scalar field in 2 space-time
dimensions. The gravitational field is now coded in a 2-metric
$\ub{g}_{ab}$ and a dilaton field $\phi$, and the theory has a
2-dimensional gravitational constant $G$ of dimensions
$[ML]^{-1}$ in addition to the constant $\k$ of dimensions
$[L]^{-1}$ ($\kappa^2$ is sometimes regarded as the
cosmological
constant).%
\footnote{In this paper we set $c=1$ but keep Newton's constant
$G$ and Planck's constant $\hbar$ free. Note that since $G\hbar$ is a
\emph{Planck number} in 2 dimensions, setting both of them to 1
is a physical restriction.}
The CGHS action is given by \cite{cghs}:
\ba \label{action2} S(\ub{g},\phi,f) = {\f{1}{G}}
&&\!\!\!\!\!\!\!\!\! {\textstyle{\int}} d^2x \sqrt{|\ub{g}|}\,
e^{-2\phi}\,(\ub{R} + 4\, \nabla^a\phi\nabla_a
\phi \nonumber\\
&+&\! 4\kappa^2) -\! {\textstyle{\f{1}{2}\int}} d^2x
\sqrt{|\ub{g}|}\, \nabla^a f \nabla_a f\, . \ea
Note that the two actions are closely related. The only difference
is in some coefficients which appear bold faced in (\ref{action1}).
This is why one expects that analysis of the CGHS model should
provide useful intuition for evaporation of spherically symmetric
black holes in 4 dimensions.

On the other hand, the two theories do differ in some important
ways and we anticipate that only certain aspects of
universality will carry over to 4-dimensions. These differences
are discussed in section \ref{s4}. Here, we only note one:
{since the dilaton field does not appear in the scalar field
action of (\ref{action2}), dynamics of $f$ decouples from that
of the dilaton.} Now, since our space-time is topologically
$\R^2$, the physical 2-metric $\ub{g}_{ab}$ is conformally
flat. We can thus fix a fiducial flat 2-metric $\eta^{ab}$ and
write $\ub{g}^{ab} = \ul{\Omega} \eta^{ab}$, thereby encoding
the physical geometry in the conformal factor $\ul{\Omega}$ and
the dilaton field $\phi$. Next, since the wave equation is
conformally invariant,
\be \label{wave} \Box_{(\ub{g})} f =0 \quad \Leftrightarrow \quad
\Box_{(\eta)} f =0\, , \ee
$f$ is only subject to the wave equation in the fiducial flat
space which can be easily solved, without any knowledge of the
physical geometry governed by $(\ul{\Omega}, \phi)$. \emph{This
is a key simplification which is not shared by the scalar field
$f$ in the spherically symmetric gravitational collapse
described by (\ref{action1}).} Denote by $z^\pm$ the advanced
and retarded null coordinates of $\eta$ so that $\eta_{ab} = 2
\partial_{(a}z^+\, \partial_{b)}z^-$. Then a general solution to
(\ref{wave}) on the fiducial Minkowski space $(M^o,\eta)$ is
simply
\be f(z^\pm) = f_+(z^+) + f_-(z^-) \ee
where $f_\pm$ are arbitrary well behaved functions of their
arguments. In the classical CGHS theory, one sets $f_- =0$ and
focuses on the gravitational collapse of the left moving mode
$f_+$. As one might expect, the true degree of freedom lies
only in $f_+$, i.e., $f_+$ completely determines the geometry.
But in the classical CGHS model, there is a further unexpected
simplification: \emph{the full solution can be expressed as an
explicit integral involving } $f_+$!

\begin{figure}
\begin{center}
1\includegraphics[width=2.8in,angle=0]{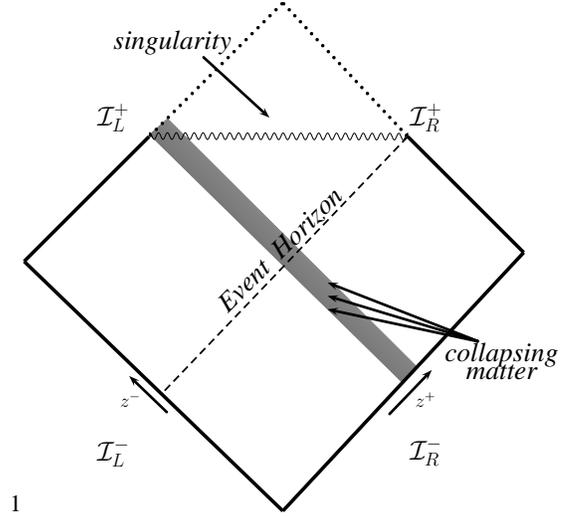}
\caption{Penrose diagram of the CGHS black hole formed by the
gravitational collapse of a left moving field $f_+$.
The physical space-time is that part of the fiducial Minkowski space
which is to the past of the space-like singularity.}
\label{BH_background}
\end{center}
\end{figure}

For later purposes, following \cite{atv}, let us set
\be \ul\Phi := e^{-2\phi}\nonumber\ee
and introduce a new field $\ul\Theta$ via
\be \ul\Theta = \ul\Omega^{-1} \ul\Phi \quad {\hbox{\rm so
that}} \quad g^{ab} = \ul{\Theta}^{-1}\, \ul\Phi\, \eta^{ab}
\nonumber\ee
Then the geometry is completely determined by the pair of
fields $\ul\Theta, \ul\Phi$, and field equations obtained by
varying (\ref{action2}) can be solved to express $\ul\Theta,
\ul\Phi$ directly in terms of $f_+$. The resulting expressions
for $\ul\Theta$ and $\ul\Phi$ are simpler in terms of
`Kruskal-like' coordinates $x^\pm$ given by
\be \k x^+ = e^{\k z^+}, \quad {\rm and} \quad \k x^- = -e^{-\k
z^-}\, .\ee
Given any regular $f_+$, the full solution to the classical CGHS
equations can now be written as
\ba \label{sol1} \ul\Theta &=& - \k^2 x^+\,x^-\nonumber\\
\ul\Phi &=& \ul\Theta - \f{G}{2}\textstyle{\int_0^{x^+}}
\dd\bar{x}^+\, \textstyle{\int_0^{\bar{x}^+}} \dd
\bar{\bar{x}}^+\,\, (\partial f_+/\partial \bar{\bar{x}}^+)^2\,
. \ea
Note that, given any regular $f$,  the fields $(\ul\Theta,
\ul\Phi)$ of (\ref{sol1}) that determine geometry are also
regular everywhere on the fiducial Minkowski manifold $M^o$.

\emph{How can the solution then represent a black hole?} It
turns out that, for any regular $f_+$, the field $\ul\Phi$ of
(\ref{sol1}) vanishes along a space-like line $\ell_s$. Along
$\ell_s$ then, $\ub{g}^{ab}$ vanishes, whence the covariant
metric $\ub{g}_{ab}$ fails to be well-defined. It is easy to
verify that the Ricci scalar of $\ub{g}_{ab}$ diverges there.
This is the singularity of the physical metric $\ub{g}$.
\emph{The physical space-time $(M, \ub{g}_{ab})$ occupies only
that portion of $M^o$ which is to the past of this singularity}
(see Fig.~\ref{BH_background}).

But does $\ell_s$ represent a \emph{black hole} singularity? It is
easy to check that $(M, \ub{g}_{ab})$ admits a smooth null infinity
$\mathcal{I}$ which has 4 components (because we are in 2 space-time
dimensions): $\mathcal{I}^-_{\rm L}$ and $\mathcal{I}^-_{\rm R}$
coincide with the corresponding $\szml$ and $\szmr$ of Minkowski
space-time $(M^o, \eta)$ while $\mathcal{I}^+_{\rm L}$ and
$\mathcal{I}^+_{\rm R}$ are proper subsets of the Minkowskian
$\szpl$ and $\szpr$. Nonetheless, $\spr$ \emph{is complete with
respect to the physical metric $\ub{g}_{ab}$ and its past does not
cover all of $M$.} Thus, there is indeed an event horizon with
respect to $\spr$ hiding a black hole singularity. However,
unfortunately $\spl$ is \emph{not} complete with respect to
$\ub{g}_{ab}$. Therefore, strictly speaking we cannot even ask%
\footnote{Even in 4-dimensions, the black hole region is defined as
$\mathcal{B} := M\, \setminus\, J^-(\mathcal{I}^+)$
\emph{provided $\mathcal{I}^+$ is complete}. If we drop the
completeness requirement, even Minkowski space would admit a
black hole! See, e.g., \cite{gh}.}
if there is an event horizon ---and hence a black hole---  with
respect to $\spl$! Fortunately, it turns out that for the analysis
of black hole evaporation ---and indeed for the issue of information
loss in full quantum theory--- only $\spr$ is relevant. \emph{To
summarize then, even though our fundamental mathematical fields
$(\ul\Theta, \ul\Phi)$ are everywhere regular on full $M^o$, a black hole
emerges because physics is determined by the Lorentzian geometry of
$\ub{g}$.}

Although a black hole does result from gravitational collapse
in the CGHS model, it follows from the explicit solution
(\ref{sol1}) that one does not encounter all the rich behavior
associated with spherical collapse in 4 dimensions. In
particular there are no critical phenomena~\cite{mc,gmg},
essentially because there is no threshold of black hole
formation: a black hole results no matter how weak the
infalling pulse $f_+$ is. However, as remarked in the beginning
of this section, the situation becomes more interesting even in
this simple model once one allows for quantum evaporation and
takes into account its back reaction.

\subsection{The Semi-Classical CGHS Model}\label{semi_intro}

To incorporate back reaction, one can use semi-classical gravity
where matter fields are allowed to be quantum but geometry is kept
classical. In this paper we will implement this idea using the mean
field approximation of \cite{atv,vat} where one ignores the quantum
fluctuations of geometry ---i.e., of quantum fields $(\hat\Theta,
\hat\Phi)$--- but keeps track of the quantum fluctuations of matter
fields. The validity of this approximation requires a large number
of matter fields $\hat{f}_i$, with $i = 1,\ldots N$ (whence it is
essentially the large $N$ approximation \cite{cghs,reviews}). Then,
there is a large domain in space-time where quantum fluctuations of
matter can dominate over those of geometry. Back reaction of the
quantum radiation modifies classical equations with terms
proportional to $N G\hbar$. However, dynamics of the physical metric
$g$ is again governed by PDEs on classical fields, $(\Theta, \Phi)$,
which we write without an under-bar to differentiate them from
solutions $(\ul\Theta, \ul\Phi)$ to the classical equations
($N\hbar=0$). In the domain of applicability of the mean-field
approximation, they are given by expectation values of the quantum
operator fields: $\Theta= \langle \h{\Theta}\rangle$ and $\Phi =
\langle \h{\Phi}\rangle$. The difference from the classical case is
that the coefficients of the PDEs and components of the metric
${g}_{ab}$ now contain $\hbar$.

\begin{figure}
\begin{center}
\includegraphics[width=2.8in,angle=0]{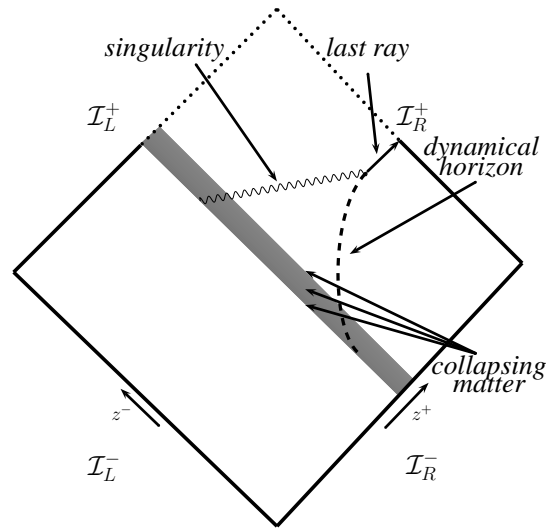}
\caption{Penrose diagram of an evaporating CGHS black hole in the mean
field approximation. Because of quantum radiation the singularity now ends
in the space-time interior and does not reach $\spl$ or $\spr$(compare with
Fig.~\ref{BH_background}.) Space-time admits a generalized dynamical horizon
whose area steadily decreases. It meets the singularity at its (right) end
point. The physical space-time in this approximation excludes a future
portion of the fiducial Minkowski space (bounded by the singularity, the last ray
and the future part of the collapsing matter).}
 \label{fethi-sketch}
\end{center}
\end{figure}

For any given finite $N$, there is nonetheless a region in which the
quantum fluctuations of geometry are simply too large for the mean
field approximation to hold. This is reflected in the fact that a
singularity persists in this approximation, although it is now
weakened. The field $\Phi$ now assumes a \emph{non-zero} value $N
G\hbar/12$ at the singularity whence ${g}^{ab}$ is invertible there
\cite{reviews}. Furthermore it is $C^0$ across this singularity but
not $C^1$. Finally, because of back-reaction, the strength of the
singularity diminishes as the black hole evaporates and the
singularity ends in the interior of space-time; in contrast to the
classical singularity, it does not reach $\spr$ (see
Fig.~\ref{fethi-sketch}). It is the dynamics of ${g}_{ab}$ that
exhibit novel, universal features.

It turns out that the fundamental equations of the mean-field
theory (and in fact also of the full quantum theory
\cite{atv,vat}) admit a \emph{scaling symmetry}, which was
discovered independently by Ori ~\cite{Ori:2010nn}. This has
important consequences, which to our knowledge have not been
fully appreciated before.%
\footnote{ In~\cite{ps} for example it was noted that $N$ could be
``scaled out'' of the problem and that the results are
``qualitatively independent of N''. However, as we will discuss in
Sec.~\ref{s2.2}, it is actually the ratio $M_{\ADM}/N$ which
classifies a solution, and more importantly there is qualitatively
different behavior in the small (Planck size) vs. large
(macroscopic) $M_{\ADM}/N$ limits. In this terminology, the
simulation of \cite{ps} corresponds to an initially Planck size
black hole.}
It naturally leads to certain physical quantities that have
universal properties during and at the end of the quantum
evaporation of large black holes. Unraveling this unforeseen
behavior requires the use of an appropriate analytical
formulation of the problem, and high precision numerical
solutions (both in terms of requiring small truncation error
and using the full range of 16-digit double precision floating
point arithmetic) ---for details, see the companion papers
\cite{rp,vat}. These universal properties have their origin in
the PDEs governing the dynamics. Therefore, our numerical
results lead to a series of interesting conjectures in
mathematical relativity and geometric analysis (discussed in
section \ref{s4}). In addition our detailed analysis shows,
quite surprisingly, that several of the standard assumptions
that have driven the quantum evaporation paradigm are
incorrect. As a consequence, results of this paper also play a
significant role in resolving the information loss puzzle in
the quantum theory \cite{atv,vat}.

\subsection{Summary of New Results}\label{summary_intro}

To summarize the new results, let us first recall the standard
paradigm. Literature on the quantum evaporation of CGHS black holes
uses a certain definition of Bondi mass $\mbt$. Essentially every
preceding paper assumed that: i) The semi-classical approximation is
excellent until the horizon shrinks to Planck size; ii) Throughout
this long phase, $\mbt$ is non-negative and the process is
quasi-static;  iii) Consequently, during this phase the quantum flux
at $\spr$ is given by the Hawking thermal flux and the
semi-classical approximation holds; and iv) At the end of this phase
the Bondi mass is also of Planck size. It is then difficult to
imagine how purity of the incoming quantum state could be preserved
in the outgoing state. However, our results show that several
features of this scenario fail to be borne out by detailed
calculations in the semi-classical theory. In particular, in section
\ref{s3} we will show the following results for a prompt collapse of
data with large ADM mass:
\begin{itemize}
\item The traditional Bondi mass, $\mbt$, in fact becomes
    \emph{negative} (and large) even while the horizon area is
    macroscopic.

\item The definition of $\mbt$ is taken directly from the
    classical theory where the black hole is static. Now, in
    4-dimensions one knows \cite{bondi} that the formula for the
    Bondi mass has to be modified in non-stationary space-times
    (from $\oint \Psi_2^o \,\dd^2V$ to $\oint (\Psi_2^o -
    \sigma\dot{\bar\sigma})\, \dd^2V$). Indeed if one were to
    ignore this modification, one would find that neither the
    Bondi mass nor the Bondi flux is always positive. By
    mimicking the 4-d procedure, Ashtekar, Taveras and
    Varadarajan (ATV)~\cite{atv} had proposed a quantum corrected
    Bondi mass, $\mbatv$, in the CGHS mean-field theory (which,
    in particular, reduces to $\mbt$ in the classical theory).
    This mass remains \emph{positive} throughout the evaporation
    process of the mean-field approximation.

\item Although the horizon area goes to zero at the end
    of the evaporation process in the mean-field approximation,
    $\mbatv$ is {\em not} of Planck size at that time (i.e., at
    the point where the `last ray' of Fig.\ref{fethi-sketch}
    intersects $\spr$). For all black holes with large initial
    ADM mass, as the horizon area shrinks to zero $\mbatv$
    approaches a \emph{universal} value $\approx 0.864 \N$ in Planck
    units, with $\N= N/24$. This end point Bondi mass is
    macroscopic since $N$ is necessarily large in the
    semi-classical theory.

\item Dynamics \emph{during} the evolution process also shows a
    \emph{universal behavior}. For example, one can calculate
    $\mbatv$ as a function of the horizon area $\a_{\rm hor}$.
    There is a transient phase immediately after the horizon is
    first formed, though after that the plot of $\mbatv$ against
    $\a_{\rm hor}$ joins a universal curve all the way to zero area.

\item The flux of energy radiated across $\spr$ departs from the
    thermal flux when $\mbatv$ and even $\a_{\rm hor}$ are
    macroscopic.

\item Although the Ricci scalar of the mean-field metric
    $g$ diverges at the (weak) singularity, it is regular
    on the last ray and goes to zero as one approaches
    $\spr$ along this ray. Thus, contrary to a wide spread
    belief, there is no `thunderbolt' curvature singularity
    in the semi-classical theory.

\end{itemize}

We will see in section \ref{s4.3} that our results strongly suggest
that the $S$ matrix from $\sml$ to $\spr$ is likely to be unitary.
However, \emph{because of the universality} of physical quantities
at $\spr$, it is very unlikely that information in the infalling
matter at $\smr$ will be recovered in the outgoing state at $\spr$.
This is in sharp contrast with a wide-spread expectation; indeed,
mechanisms for information recovery have been suggested in the past
(see e.g. \cite{st}). This expectation illustrate the degree to
which universality was unanticipated in much of the CGHS literature.

The rest of this paper is organized as follows. In section \ref{s2}
we summarize the mean-field theory of \cite{atv,vat}, introduce the
scaling behavior and explain the subtleties associated with Planck
units in 2 dimensions. This framework provides the basis for the
numerical results discussed in \ref{s3}. These results are more
extensive than the brief summary presented above. Finally in section
\ref{s4} we discuss the ramifications of these results and list
interesting problems they suggest in geometric analysis.

\section{The mean-field Approximation}
\label{s2}

This section is divided into three parts. We recall (mainly from
\cite{reviews,atv,vat}) the PDEs that govern semi-classical gravity
in the first part and their consequences in the second. In the third
we introduce a scaling symmetry which leads to a natural distinction
between macroscopic and Planck scale black holes.

\subsection{mean-field equations}
\label{s2.1}

As mentioned in section \ref{s1}, in semi-classical gravity we
have $N$ quantum scalar fields $\h{f}^{(i)}$, with $i = 1\ldots
N$. In the mean-field approximation, we capture the idea that
it is only the left moving modes that undergo gravitational
collapse by choosing the initial state appropriately: we let
this state be the vacuum state for the right moving modes
$\hat{f}_-^{(i)}$ and a coherent state peaked at any given
classical profile $f^o_+$ for \emph{each} of the\, $N$\, left
moving fields $\hat{f}_+^{(i)}$. This specification at
$\mathcal{I}^-$ defines a (Heisenberg) state $|\Psi\rangle$.
Dynamical equations are obtained by taking expectation values
of the quantum evolution equations for (Heisenberg) fields in
this state $|\Psi\rangle$ and ignoring quantum fluctuations of
geometry but not of matter. Technically, this is accomplished
by substituting polynomials $P(\h\Theta, \h\Phi)$ in the
geometrical operators with polynomials $P(\langle
\h\Theta\rangle,\,\langle \h\Phi\rangle) := P(\Phi,\Theta)$
of their expectation values. For the matter fields
$\h{f}^{(i)}$, on the other hand, one does not make this
substitution; one keeps track of the quantum fluctuations of
matter. These lead to a conformal anomaly: While the trace of
the stress-tensor of scalar fields vanishes in the classical
theory due to conformal invariance, the expectation value of
this trace now fails to vanish. Therefore equations of motion
of the geometry acquire new source terms of quantum origin which
modify its evolution.

To summarize, then, in the mean-field approximation the dynamical
objects are again just smooth fields $f_i, \Theta,\Phi$
(representing expectation values of the corresponding quantum
fields). While there are $N$ matter fields, geometry is still
encoded in the two basic fields $\Theta,\Phi$ which determine the
space-time metric $g^{ab}$ via $g^{ab} = \Omega \eta^{ab} :=
\Theta^{-1}\, \Phi\, \eta^{ab}$. Dynamics of $f^{(i)}, \Theta, \Phi$
are again governed by PDEs but, because of the trace anomaly,
equations governing $\Theta, \Phi$ acquire quantum corrections which
encode the back reaction of quantum radiation on geometry.

As in 4-dimensional general relativity, there are two sets of
PDEs: Evolution equations and constraints which are preserved
in time. As in the classical theory, it is simplest to fix the
gauge and write these equations using the advanced and retarded
coordinates $z^\pm$ of the fiducial Minkowski metric. The
evolution equations are given by
\be \label{de1}\Box_{(\eta)}\, f^{(i)} = 0\quad \Leftrightarrow
\quad\Box_{(g)}f^{(i)} = 0.\ee
for matter fields and
\ba \label{de2}\Dp\,\Dm\, \Phi + \k^2 \Theta = G\,
\langle\h{T}_{+-}\rangle  &\equiv& {\N G\hbar}\, \Dp\, \Dm\, {\ln
\Phi\Theta^{-1}}\\
\label{de3} \Phi \Dp\,\Dm {\ln \Theta} =  - G\,
\langle\h{T}_{+-}\rangle &\equiv&  -{\N G\hbar}\, \Dp\, \Dm\, {\ln
\Phi\Theta^{-1}}\nonumber\\ \ea
for the geometrical fields where, as before, $\N = N/24$. The
constraint equations tie the geometrical fields $\Theta,\Phi$ to the
matter fields $f_i$. They are preserved in time. Therefore we can
impose them just at $\mathcal{I}^-$ where they take the form:
\be \label{ce1}  -\Dm^2\, \Phi + \Dm\,\Phi \Dm\, \ln \Theta \,=\,
G\,\langle \h{T}_{- -}\rangle\,\, \h{=}\,\,0\ee
and
 \be \label{ce2} -\Dp^2\, \Phi + \Dp\,\Phi \Dp\, \ln \Theta
\,=\, G\, \langle \h{T}_{++}\rangle\,\h{=}\, 12\N G\, (\Dp
f_+^{o})^2 \ee
where $\h{=}$ stands for `equality at $\mathcal{I}^-$.'

We will conclude this discussion of the field equations with a few
remarks, and a description of our initial conditions.
Because $\h{f}_-^{(i)}$ are all in the vacuum state, it
follows immediately that, as in the classical theory, all the right
moving fields vanish; $f_-^{(i)} =0 $ also in the mean-field theory.
Similarly, because $\h{f}_+^{(i)}$ are in a coherent state peaked at
some classical profile $f^o_+$, it follows that, for all $i$,
$f_+^{(i)} (z^+) = f_+^o(z^+)$ (on the entire fiducial Minkowski
manifold $M^o$). Thus, as far as matter fields are concerned, there
is no difference between the classical and mean-field theory.
Similarly, as in the classical theory, we can integrate the
constraint equations to obtain initial data on two null
hypersurfaces. We will assume that $f^{(o)}_+$ vanishes to the past
of the line $z^+=z^+_o$. Let $I^-_{\rm L}$ denote the line $z^+ =
z^+_o$ and $I^-_{\rm R}$ the portion of the line $z^- = z^-_o \ll
- 1/\k$ to the future of $z^+ = z^+_o$. We will specify initial
data on these two surfaces. The solution to the constraint equations
along these lines is not unique and, as in the classical theory we
require additional physical input to select one. We will again
require that $\Phi$ be in the dilaton vacuum to the past of
$I^-_{\rm L}$ and by continuity on $I^-_{\rm L}$. Following the CGHS
literature, we will take it to be $\Phi = e^{\k(z^+ -z^-)}$.
\footnote{\label{fn5} Strictly, since $\h\Phi$ is an operator on the tensor
product of $N$ Fock spaces, one for each $\h{f}^{(i)}$, the
expectation value is $N e^{\k(z^+ -z^-)}$. But this difference
can be compensated by shifting $z^-$. We have chosen to use the
convention in the literature so as to make translation between
our expressions and those in other papers easier.}
Thus, the initial values of semi-classical $\Theta,\Phi$
coincide with those of classical $\ul\Theta, \ul\Phi$:
\be \label{sol2} \Theta\, \h{=} \, e^{\k(z^+_o -z^-)} \quad
\hbox{\rm on all of $I^-_{\rm L}$ and $I^-_{\rm R}$} \ee
and
\ba \label{sol3} \Phi \, &\h{=}&\, \Theta \quad \hbox{\rm
on}\,\,\, I^-_{\rm L}  \,\,\, {\rm and,}\nonumber\\
\Phi\, &\h{=}&\, \Theta -
12\bar{N}G\textstyle{\int_{-\infty}^{z^+}} \dd\bar{z}^+\,
e^{\k\b{z}^+}\, \textstyle{\int_{-\infty}^{\bar{z}^+}} \dd
\bar{\bar{z}}^+\,\, e^{-\k \b{\b{z}}^{+}}\,(\f{\partial
f_+^{(o)}}{\partial \bar{\bar{z}}^+})^2\nonumber\\
&&{}\quad\quad\quad\hbox{\rm on $I^-_{\rm R}$} \ea
(see (\ref{sol1})). The difference in the classical and
semi-classical theories lies entirely in the evolution equations
(\ref{de2}) and (\ref{de3}). In the classical theory, the right hand
sides of these equations vanish whence one can easily integrate
them. In the mean-field theory, this is not possible and one has to
take recourse to numerical methods. Finally, while our analytical
considerations hold for any regular profile $f_+^{o}$, to begin with
we will follow the CGHS literature in Sec.~\ref{s3.1} and
Sec.~\ref{s3.2} and specify $f_+^{o}$ to represent a collapsing
shell:
\be \label{shell} 12\bar{N}\, \left(\f{\partial f_+^{o}}{\partial
z^+}\right)^2 = M_{\ADM} \, \delta(z^+) \ee
so the shell is concentrated at $z^+ = 0$. In the literature
this profile is often expressed, using $x^+$ in place of $z^+$,
as:
\be \label{shell2} 12\bar{N}\, \left(\f{\partial \t{f}_+^{o}}{\partial
x^+}\right)^2 = M_{\ADM} \, \delta(x^+ - \f{1}{\k}) \ee
where $\t{f}^{(o)}(x^+) = f^{(o)}(z^+)$. In Sec.~\ref{s3.3} we
will discuss results from a class of smooth matter profiles.

\subsection{Singularity, horizons and the Bondi mass}
\label{s2.2}

The classical solution (\ref{sol1}) has a singularity $\ell_s$ where
$\Phi$ vanishes. As remarked in section \ref{s1}, in the mean-field
theory, a singularity persists but it is shifted to $\Phi = 2\N
G\hbar$ \cite{reviews}. The metric $g^{ab} = \Theta^{-1} \Phi\,
\eta^{ab}$ is invertible and continuous there but not $C^1$. Thus
the singularity is weakened relative to the classical theory.
Furthermore, its spatial extension is diminished. As indicated in
Fig.\ref{fethi-sketch}, the singularity now originates at a finite
point on the collapsing shell (i.e. does not extend to $\spl$) and
it ends in the space-time interior (i.e., does not extend to
$\spr$).

What is the situation with horizons? Recall from section
\ref{s1} that, in the spherically symmetric reduction from
4-dimensions, $r^2 = e^{-2\phi}/\k^2 := \Phi/\kappa^2$ and each
round 2-sphere in 4-dimensional space-time projects down to a
single point on the 2-manifold $M$. Thus, in the CGHS model we
can think of $\Phi$ as defining the `area' associated with any
point. (It is dimensionless because in $D$ space-time
dimensions the area of spatial spheres has dimensions
$[L]^{D-2}$.) Therefore  it is natural to define a notion of
trapped points: A point in the CGHS space-time $(M,g)$ is said
to be \emph{future trapped} if $\Dp \Phi$ and $\Dm \Phi$ are
both negative there and \emph{future marginally trapped} if
$\Dp \Phi$ vanishes and $\Dm \Phi$ is negative there
\cite{reviews,sh}. In the classical solution resulting from the
collapse of a shell (\ref{shell}), all the marginally trapped
points lie on the event horizon and their area is a constant;
we only encounter an isolated horizon \cite{akrev} (see
Fig.\ref{BH_background}). The mean-field theory is much richer
because it incorporates the back reaction of quantum radiation.
In the case of a shell collapse, the field equations now imply that
a marginally trapped point first forms at a point on the shell
and has area
\ba \label{dh} \a_{\rm initial} &:=& (\Phi - 2\N G\hbar)|_{\rm
initial}
\nonumber\\
&=&  -\N G\hbar + \N G\hbar\, \left(1
+\f{M_{\ADM}^2}{\N^2\hbar^2\k^2} \right)^{\textstyle\f{1}{2}} \ea
As time evolves, this area \emph{shrinks} because of quantum
radiation \cite{reviews}. The world-line of these marginally trapped
points forms a \emph{generalized dynamical horizon} (GDH),
`generalized' because the world-line is time-like rather than
space-like \cite{akrev}. (In 4 dimensions these are called
marginally trapped tubes \cite{ak}.) The area finally shrinks to
zero. This is the point at which the GDH meets the end-point of the
(weak) singularity \cite{ps,dl,reviews} (see
Fig.\ref{fethi-sketch}). It is remarkable that all these interesting
dynamics occur simply because, unlike in the classical theory, the
right sides of the dynamical equations (\ref{de2}), (\ref{de3}) are
non-zero, given by the trace-anomaly.

We will see in section \ref{s3} that while the solution is
indeed asymptotically flat at $\spr$, in contrast to the
classical solution, $\spr$ \emph{is no longer complete.} More
precisely, the space-time $(M,g)$ now has a future boundary at
the last ray ---the null line to $\spr$ from the point at which
the singularity ends--- and the affine parameter along $\spr$
\emph{with respect to $g_{ab}$ has a finite value} at the point
where the last ray meets $\spr$. Therefore, in the
semi-classical theory, \emph{we cannot even ask if this
space-time admits an event horizon.} While the notion of an
event horizon is global and teleological, the notion of trapped
surfaces and GDHs is quasi-local. As we have just argued, these
continue to be meaningful in the semi-classical theory. What
forms and evaporates is the GDH.

Next, let us discuss the structure at null infinity
\cite{atv,vat}. As in the classical theory, we assume that the
semi-classical space-time is asymptotically flat at $\spr$ in
the sense that, as one takes the limit $z^+ \rightarrow \infty$
along the lines $z^- = {\rm const}$, the fields $\Phi, \Theta$
have the following behavior:
\ba  {\Phi} &=& A(z^-)\, e^{\k z^+} + B(z^-) +
O(e^{-\k z^+}) \nonumber\\
{\Theta} &=& \ub{A}(z^-)\, e^{\k z^+} + \ub{B}(z^-) + O(e^{-\k
z^+})\, , \label{asym} \ea
where $A, B, \ub{A}, \ub{B}$ are some smooth functions of
$z^-$. Note that the leading order behavior in (\ref{asym}) is
the same as that in the classical solution. The only difference
is that $B, \ub{B}$ are not required to be constant along
$\spr$ because, in contrast to its classical counterpart, the
semi-classical space-time is non-stationary near null infinity
due to quantum radiation. Therefore, as in the classical
theory, $\spr$ can be obtained by taking the limit $z^+
\rightarrow \infty$ along the lines $z^- = {\rm const}$. The
asymptotic conditions (\ref{asym}) on $\Theta, \Phi$ imply that
curvature ---i.e., the Ricci scalar of $g_{ab}$--- goes to zero
at $\spr$. We will see in section \ref{s3} that these
conditions are indeed satisfied in semi-classical space-times
that result from collapse of matter from $\smr$.

Given this asymptotic fall-off, the field equations determine
$\ub{A}$ and $\ub{B}$ in terms of $A$ and $B$. The metric $g_{ab}$
admits an \emph{asymptotic} time translation $t^a$ which is unique
up to a constant rescaling and the rescaling freedom can be
eliminated by requiring that it be (asymptotically) unit. The
function $A(z^-)$ determines the affine parameter $y^-$ of $t^a$
via:
\be  \label{yz} e^{-\k y^-} =\, A(z^-). \ee
Thus $y^-$ can be regarded as the unique asymptotic time parameter
with respect to $g_{ab}$ (up to an additive constant). Near $\spr$
the mean-field metric ${g}$ can be expanded as:
\be \dd {S}^2 = - \left(1 + {B} e^{\k (y^- - y^+)} + O(e^{-2\k
y^+})\right)\,\dd y^+ \, \dd y^- \ee
where $y^+ = z^+$.

Finally, equations of the mean-field theory imply \cite{atv,vat}
that there is a balance law at $\spr$:
\ba \label{balance}\f{\dd}{\dd y^-}\big[ \f{\dd{B}}{\dd y^-} &+& \k
{B}\, +\, {\N \hbar G}\, \big(\f{\dd^2 y^-}{\dd z^{-2}}\, (\f{\dd
y^-}{\dd z^-})^{-2}\,
\big)\,\big] \nonumber\\
&=& - \f{\N \hbar G}{2}\, \big[\f{\dd^2 y^-}{\dd z^{-2}}\, (\f{\dd
y^-}{\dd z^-})^{-2}\,\,\big]^2\, . \ea
In \cite{atv}, this balance law was used to introduce a new notion
of Bondi mass and flux. The left side of (\ref{balance}) led to the
definition of the Bondi mass:
\be \label{mass}\mbatv = \f{\dd{B}}{\dd y^-} + \k {B}\, +\, {\N
\hbar G}\, \big(\f{\dd^2 y^-}{\dd z^{-2}}\, (\f{\dd y^-}{\dd
z^-})^{-2}\, \big)\, ,\ee
while the right side provided the Bondi flux:
\be\label{flux}  \fatv =  \f{\N \hbar G}{2}\, \big[\f{\dd^2
y^-}{\dd z^{-2}}\, (\f{\dd y^-}{\dd z^-})^{-2}\,\,\big]^2\,
,\ee
so that we have:
\be \f{\dd \mbatv}{\dd y^-} = - \fatv\, . \ee
By construction, as in 4 dimensions, the flux is manifestly
positive so that $\mbatv$ decreases in time. Furthermore, it
vanishes on an open region if and only if $y^- = C_1 z^- +
C_2$ for some constants $C_1, C_2$, i.e. if and only if the
asymptotic time translations defined by the physical, mean
field metric $g$ and by the fiducial metric $\eta$ agree at
$\spr$, or, equivalently, \emph{if and only if the asymptotic
time translations of $g$ on $\sml$ and $\spr$ agree}. Finally,
note that ${g}^{ab} = \eta^{ab},\, {f}_\pm =0,\,  \Phi = \Theta
= \exp \k(z^+-z^-)$, is a solution to the full mean-field
equations. As one would expect, both $\mbatv$ and $\fatv$
vanish for this solution.

The balance law is just a statement of conservation of energy. As
one would expect, $\hbar$ appears as an overall multiplicative
constant in (\ref{flux}); in the classical theory, there is no flux
of energy at $\spr$. If we set $\hbar =0$,  $\mbatv$ reduces to the
standard Bondi mass formula in the classical theory (see e.g.,
\cite{sh}). Previous literature \cite{cghs,reviews,hs,sh,tu,st,dl}
on the CGHS model used this classical expression also in the
semi-classical theory. Thus, in the notation we have introduced
here, the traditional definitions of mass and flux are given by
\be \mbt = \f{\dd{B}}{\dd y^-} + \k {B}\, ,\ee
and
\be \ftrad = \fatv + {\N \hbar G}\, \f{\dd}{\dd y^-} \big(\f{\dd^2
y^-}{\dd z^{-2}}\, (\f{\dd y^-}{\dd z^-})^{-2}\, \big)\, . \ee
As noted in section \ref{s1}, numerical simulations have shown that
$\mbt$ can become negative and large even when the horizon area is
large, while $\mbatv$ remains positive throughout the evaporation
process.

\subsection{Scaling and the Planck regime}
\label{s2.3}

Finally, we note a scaling property of the mean-field theory, which
Ori recently and independently also uncovered~\cite{Ori:2010nn} and
which is also observed in other quantum gravitational 
systems~\cite{Yeom:2009}. We
were led to it while attempting to interpret numerical results which
at first seemed very puzzling; it is thus a concrete example of the
how useful the interplay between numerical and analytical studies
can be. Let us fix $z^\pm$ and regard all fields as functions of
$z^\pm$. Consider any solution $(\Theta, \Phi, N, f_+^{(i)})$ to our
field equations, satisfying boundary conditions (\ref{sol2}) and
(\ref{sol3}). Then, given a positive number $\lambda$, $(\t\Theta,
\t\Phi, \t{N}, \t{f}^{(i)}_+)$ given by%
\footnote{The shift in $z^-$ is needed because we chose to use the
initial value $\Theta = e^\k(z^+-z^-)$ on $I^-_{\rm L}$ as in the
literature rather than $\Theta = N e^{\k(z^+-z^-)}$. See footnote
\ref{fn5}.}
\ba \label{scaling}
 \t\Theta (z^+, z^-) &=& \lambda \Theta(z^+, z^- + \f{\ln\lambda}{\k}),
\quad \t{N} = \lambda N \nonumber\\
 \t\Phi (z^+, z^-) &=& \lambda \Phi(z^+, z^- +\f{\ln\lambda}{\k}),
 \quad \t{f}^{(i)}_+(z^+) = f^{(i)}_+ (z^+)\nonumber \ea
is also a solution satisfying our boundary conditions, where, as
before, we have assumed that all scalar fields have an identical
profile $f^{o}_+$. Note that $f^{o}_+$ is completely general; we
have not restricted ourselves, e.g., to shells. Under this
transformation, we have
\ba \bar{g}^{ab}
&\rightarrow& \bar{g}^{ab}\nonumber\\
y^- &\rightarrow& y^- - \f{1}{\kappa}\ln \lambda \nonumber\\
M_{\ADM} &\rightarrow& \lambda M_{\ADM}\nonumber\\
\mbatv &\rightarrow& \lambda \mbatv\nonumber\\
\fatv &\rightarrow& \lambda \fatv\nonumber\\
\mathbf{a}_{\GDH} &\rightarrow& \lambda\, \mathbf{a}_{\GDH} \ea
where $\a_{\GDH}$ denotes the area of the generalized dynamical
horizon. This symmetry implies that, \emph{as far as space-time
geometry and  energetics are concerned, only the ratios $M/N$
matter, not separate values of $M$ and $N$ themselves} (where
$M$ can either be the ADM or the Bondi mass). Thus, for
example, whether for the evaporation process a black hole is
`macroscopic' or `Planck size' depends on the ratios $M/N$ and
$\a_{\GDH}/N$ rather than on the values of $M$ or $\a_{\GDH}$
themselves.

We will set
\ba M^\star &=& M_{\ADM}/{\N}\nonumber \\
M^\star_{\Bondi} &=& \mbatv/\N, \,\, {\rm and}\nonumber\\
m^\star &=& M^\star_{\Bondi}|_{\hbox{\rm last ray}}
\label{star}\ea
(We use ${\N} = N/24$ in these definitions because the
dynamical equations feature $\N$ rather than $N$.) We will need
to compare these quantities with the Planck mass. Now, in 2
dimensions, $G,\,\hbar$ and $c$ do not suffice to determine
Planck mass, Planck length and Planck time uniquely because
$G\hbar$ is dimensionless. But in 4 dimensions we have
unambiguous definitions of these quantities and, conceptually,
we can regard the 2 dimensional theory as obtained by its
spherical reduction. In 4-dimensions, (using the c=1 units used
here) the Planck mass is given by $M_{\rm Pl}^2 = \hbar/G_4$
and the Planck time by $\tau_{\rm Pl}^2= G_4\,\hbar$. From Eqs
(\ref{action1}) and (\ref{action2}) it follows that $G_4$ is
related to the 2-dimensional Newton's constant $G$ via $ G= G_4
\k^2$. Therefore we are led to set
\be  \label{pl} M_{\rm Pl}^2  = \f{\hbar\kappa^2}{G}, \quad{\rm and}
\quad \tau_{\rm Pl}^2 = \f{G\hbar}{\kappa^2}\, . \ee
When can we say that a black hole is macroscopic? One's first
instinct would be to say that the ADM mass should be much larger
than $M_{\rm Pl}$ in (\ref{pl}). But this is not adequate for the
evaporation process because the process depends also on the number
of fields $N$. In the external field approximation where one ignores
the back reaction, we know that at late times the black hole
radiates as a black body at a fixed temperature $T_{\rm Haw} = \k
\hbar$.
\footnote{Note that this relation is the same as that in 4
dimensions because the classical CGHS black hole is stationary
to the future of the collapsing matter with surface gravity
$\k$. However, there is also a \emph{key} difference: now $\k$
is just a constant, independent of the mass of the black hole.
Therefore, unlike in 4-dimensions, the temperature of the CGHS
black hole is a universal constant in the external field
approximation. Therefore, when back reaction is included, one
does not expect a CGHS black hole to get hotter as it shrinks.}
The Hawking energy flux at $\spr$ is given by $F^{\rm Haw} =
\N\kappa^2\hbar/2$. Therefore the evaporation process will last
much longer than 1 Planck time if and only if $(M_{\ADM}/
F^{\rm Haw}) \gg \tau_{\rm pl}$,\, or, equivalently
\be \label{macro} M^\star \gg G\hbar\,\, M_{\rm Pl} .\ee
(Recall that $G\hbar$ is the Planck number.) So, a necessary
condition for a black hole to be macroscopic is that $M^\star$
should satisfy this inequality. In section \ref{s3} we will see
that, in the mean-field theory, quantum evaporation reveals
universality already if $M^\star \gtrsim 4\,G\hbar\, M_{\rm
Pl}$. 

\section{Anticipated and Unforeseen Behavior}
\label{s3}

All physical predictions of the mean-field theory arise from
the set of 5 equations (\ref{de1}) -- (\ref{ce2}). The only
difference from the classical theory lies in the fact that,
because of the trace anomaly, right hand sides of the dynamical
equations (\ref{de2}) and (\ref{de3}) are no longer zero. But
this difference has very significant ramifications. In
particular, it is no longer possible to obtain explicit
analytical solutions; one has to take recourse to
numerics.%
\footnote{There are variants of the CGHS model that are explicitly
soluble, for example the RST (Russo-Susskind-Thorlacius)
model~\cite{rst}, and the Bilal-Callan model~\cite{Bilal:1992kv}.
However, results obtained in these models are not likely to be
generic even in 2 dimensions because of their extra symmetries
\cite{hs,ps}. More importantly, it was pointed out in
\cite{reviews,ps,Strominger:1994ey} that the RST model is
inconsistent even in the large $N$ limit, and the Bilal-Callan model
has a Hamiltonian that is unbounded from below. Thus though they
exhibit many features of general 2D semi-classical black hole
evaporation, they are physically less interesting than the CGHS
model.}

Also, we now have interesting time-dependent phenomena such as
formation and evaporation of dynamical horizons. Given these
differences, a number of global questions naturally arise. Does
the space-time continue to be asymptotically flat at $\spr$ in
the mean-field theory? If so, is $\spr$ complete as in the
classical theory? Is the Bondi mass positive? Does it go to the
Planck scale as the horizon area goes to zero? Is the flux of
energy of the quantum radiation constant, given by
$\hbar\k^2\N/2$ at late times, as in the external field
approximation {\em a~la}~Hawking \cite{swh,gn}? If not, the quantum
radiation would not be compatible with thermal flux, violating
a key assumption that has been made over the years.

Recall from section \ref{s2.2} that since our profile functions
$f^{(o)}_+$ vanish to the past of a null ray $I^-_{\rm L}$, the
solution in the past is given by $f^{(i)} =0,\, g_{ab} =
\eta_{ab},\, \Theta = \Phi = e^{\k(z^+-z^-)}$. To obtain it to
the future of $I^-_{\rm L}$, we use the initial data on
$I^-_{\rm L}$ and $I^-_{\rm R}$ given by (\ref{sol2}) and
(\ref{sol3}). The dynamical equations (\ref{de1}), (\ref{de2})
are hyperbolic and therefore well adapted to numerical
evolution. They have been solved numerically before (see, in
particular, \cite{dl,ps,hs,tu}), and much information
has been learnt about the CGHS model, for example
the dynamics of the GDH and that it evaporates to zero
area, terminating in a naked singularity. However, to our knowledge in all
these simulations, the choice of parameters $M_\ADM$ and $N$
was such that $M^\star = M_{\ADM}/\N$ was less that 2.5. As we
will see, in a precise sense, these black holes are Planck
scale already when they are formed.

Reliable simulations of macroscopic black holes with $M^\star
\gtrsim 5-10$ turn out to be much harder to perform, and several
additional steps beyond a straight-forward discretization are
necessary to study this regime \cite{rp}. First, one needs to
formulate the problem in terms of `regularized' variables which do
not diverge at infinity. Second, one needs to introduce coordinates
which (a) bring the infinite portion of $\spr$ of interest to a
compact interval, and (b) enlarge the region near the last ray by a
factor of roughly $e^{M^\star}$ relative to a uniform discretization
of the (compactified) time-translation coordinate $z^-$. (In other
words, a region in the vacuum solution near $\sml$ of physical size
equal to $\sim e^{-M^\star}$ expands to a region of physical size
$O(1)$ on $\spr$, where all the interesting dynamics occurs.) Third,
to achieve this latter part of the coordinate transformation one
needs to know the location of the last ray extremely well, requiring
high accuracy numerical methods. This is achieved by using
Richardson extrapolation with intermittent error removal, beginning
with a unigrid method that is second order accurate. With four
successively finer meshes the overall rate of convergence of the
scheme is $\mathcal{O}(h^7)$, and this was sufficient to reduce the
truncation error to the order of machine round-off ($\sim
10^{-16}$). These high precision numerical calculations showed that,
while some of the long held assumptions on the nature of quantum
evaporation are borne out, several are not. We summarize these
results from a physical perspective in the next three sub-sections.
Complementary discussion of numerical issues appears in \cite{rp}.

Numerical calculations were performed for a range of rescaled ADM
masses $M^\star$ from $2^{-10}$ to $16$ and $N$ varying from $12$ to
$24000$. Since in this paper we are primarily interested in black
holes which are initially macroscopic, we will focus on $M^\star \ge
1$ and, since our simulations exhibited the expected scaling
behavior, all our plots, except Fig.~\ref{N720M360M_B_ATV_and_Trad},
are for $N =24$ (i.e., $\N =1$). Finally, in these simulations we
set $\hbar=G=\kappa=1$. The first two sub-sections summarize results
from a shell collapse and the third sub-section reports on results
obtained from more general infalling profiles.

\subsection{Shell Collapse: Anticipated Behavior}
\label{s3.1}

\begin{figure}
\includegraphics[scale=0.62]{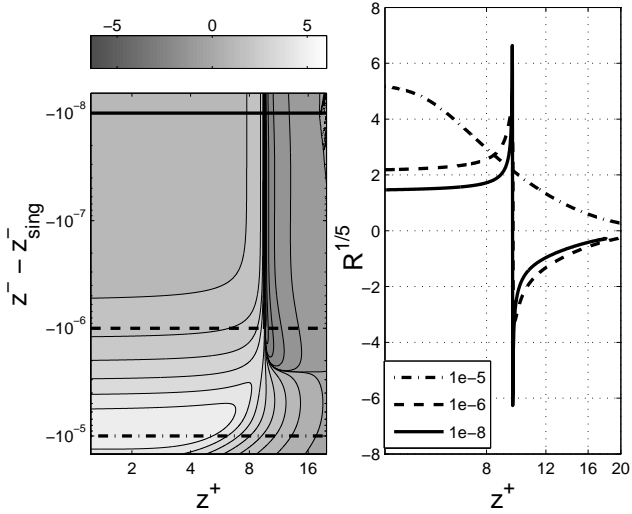}
\caption{ The Ricci scalar $R$ for $M^\star=8$. Left:
2D contour plot of $R^{1/5}$ showing the increase in $R$ as
the singularity (dark vertical region near the middle) is approached
and the asymptotically flat region ($R\to0$) near $\scri^+_R$
($z^+ \to \infty$).
Right: $R^{1/5}$ as a function of $z^+$ on the lines $z^- - z^-_{\rm sing} =
-10^{-5}, -10^{-6}, -10^{-8}$ (marked on the left panel as horizontal
lines), showing a double peak, indicating the divergent
behavior of $\partial_+ \partial_- \Phi$ at the singularity. The fact that the
peak is narrow rules out a strong thunderbolt singularity.
Note that the dark color at the region of the
singularity is due to the high density of contour lines, and not
directly due to negative values of $R$.
While naive numerical calculation of $R$ very close
to $\scri_R^+$
does not yield reliable results due to catastrophic cancelation,
it is already very
small in the high $z^+$ values shown here, and the trend towards $0$ is clear.
}
\label{N24M8Ricci}
\end{figure}
\emph{Asymptotic flatness at $\scri^+_R$}: First, $\Theta,\Phi$
do indeed satisfy the asymptotic conditions (\ref{asym}). This
was also noted in the recent approximate solution to the CGHS
equations by Ori~\cite{Ori:2010nn}. The simulations provide
values of the functions $A(z^-), B(z^-)$ and $y^-(z^-)$. As a
consistency check on the simulation, we verified the balance
law (\ref{balance}) by calculating separately the right and
left sides of this equation as close to the last ray as the
numerical solution gave reliable (convergent) results. We also
computed the scalar curvature $R$ of the mean-field metric $g$,
and it does go to zero at $\spr$---see Fig.~\ref{N24M8Ricci}
for an example.
\begin{figure}
\includegraphics[scale=0.65]{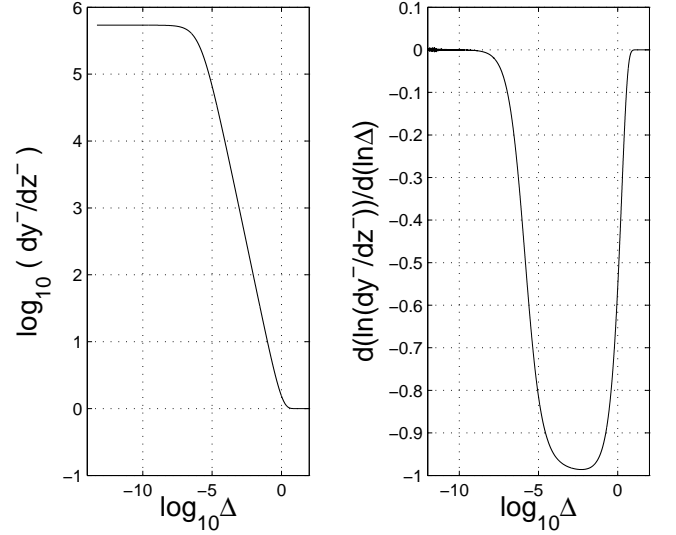}
\caption{ Left: Plot of $\log_{10} ({\dd y^-}/{\dd z^-})$\, vs\, $\log_{10}
\Delta$\,\, for $M^\star=8$, $\N=1$, where $\Delta = \left( z^-_{\rm sing} -
z^- \right)$. Right: Slope of the curve on the left.
If locally the function on the left behaves as $\sim (\k \Delta)^{-p}$,
the curve on the right shows $-p$.
In the distant past
(rightmost region in both plots), $y^-$ tends to $z^-$. The intermediate
region is similar to that in the classical solution where $({\dd y^-}/{\dd z^-})
\sim (\k \Delta)^{-1}$. As the last ray is further approached
(leftmost region), we see that $({\dd y^-}/{\dd z^-})$ increases much slower
than $(\k \Delta)^{-1}$, leading to a finite value for $y^-$ at the last ray.
}
\label{N24M8dydz_vs_z_logarithmic}
\end{figure}
\medskip

\emph{Finiteness of $y^-$ at the last ray}:  In the classical
solution, the affine parameters $\ub{y}^-$ along $\spr$ and $z^-$
along $\sml$ defined by $\ub{g}$ are related by
\begin{equation} \label{uby}
e^{-\kappa \ub{y}^-} = e^{-\kappa z^-} - \textstyle{\frac{GM}{\kappa}}
\, .
\end{equation}
Hence $\ub{y}^- =\infty$ at $\k z^- = - \ln (GM/\k)$. This is the
point at which the singularity and the event horizon meet $\spr$
(see Fig \ref{BH_background}). Thus, in the classical solution
$\spr$ is complete but, in a precise sense, smaller than $\sml$. For
a test quantum field $\h{f}_-$ on the classical solution, one then
has to trace over modes residing on the part of $\sml$ which
is missing from $\spr$. This fact is directly responsible for pure
states on $\sml$ to evolve to mixed states on $\spr$, i.e., for the
non-unitarity of the $S$-matrix \cite{atv,vat} of the test field.
What is the situation in the mean-field theory? Our analysis shows
that, as generally expected, the affine parameter w.r.t. the mean
field metric $g$ takes a \emph{finite} value at the last ray on
$\spr$; a necessary condition for unitarity of the S-matrix is met.

To establish this result, we apply the following strategy. Let us
return to the classical solution $\ub{g}$ and set
\be \k z_{\rm sing,cl}^- = -\ln ({GM}/{\k}) \quad {\rm and}\quad
\Delta_{\rm cl} = z_{\rm sing,cl}^-\, - z^- \, .\ee
(The subscript `sing,cl' just highlights the fact that this is
the point at which the classical singularity meets $\spr$.)
Then we have
\begin{equation}
\ub{y}^- = z^- -\frac{1}{\kappa} \ln \left( 1 - e^{-\kappa \Delta_{\rm cl}}
\right) \ .
\end{equation}
When $\Delta_{\rm cl}$ tends to zero, $\ub{y}^-$ is dominated by the
leading order term  $-(1/\k)\, \ln(-\k \Delta_{\rm cl})$ which
diverges at $\Delta_{\rm cl}=0$. This logarithmic divergence is
coded in the power $-1$ in the expression of the derivative $(\dd
\ub{y}^-/\dd z^-)$:
\begin{equation}
\frac{\dd \ub{y}^-}{\dd z^-} = (\k\, \Delta_{\rm cl})^{-1}
 + \textrm{finite terms} \ .
\end{equation}
If we had $(\k\, \Delta_{\rm cl})^{-p}$ on the right side rather
than $(\k\,\Delta_{\rm cl})^{-1}$, then $\ub{y}^-$ would have been finite
at the future end of $\spr$ of $\ub{g}$ for $p<1$ (as then $\ub{y}^-
= (\k\, \Delta_{\rm cl})^{1-p}/(1-p)+$ \textrm{ finite terms}).

In the mean-field theory, the last ray starts at the end point of
the singularity and meets $\spr$ of $g$ at its future end point. We
will denote it by the line $z^- = z^-_{\rm sing}$. Following
the above discussion, to show that the
affine parameter $y^-$ w.r.t. $g$ is finite at $z^-= z^-_{\rm
sing}$ we focus on the behavior of $(\dd y^-/\dd
z^-)$ near this future end point of $\spr$. More precisely, we
analyze the functional behavior of $({\dd y^-}/{\dd z^-})$ and
determine a local $p$ extracted from the logarithmic derivative of
$({\dd y^-}/{\dd z^-})$ with respect to $\Delta \equiv z_{\rm
sing}^- -z^-$. Results in Fig.~\ref{N24M8dydz_vs_z_logarithmic} show
that $({\dd y^-}/{\dd z^-})$ grows much slower near the last ray in
the mean-field theory than it does in the classical theory.
In fact, over the entire range of $\spr$
the local estimate of $p$ is strictly less that $1$,
and asymptotes to $0$ approaching the last ray.
This implies that $y^-$ is finite at the last ray in the mean-field
theory.

Note that the above analysis is only valid if we have determined
the location of the singularity with sufficient accuracy such
that the numerical uncertainty in its location is
much smaller than the range in $\Delta$ where
we extract the asymptotic behavior of the function.
From convergence studies we estimate our precision in determining $z^-_{\rm sing}$
is at the order of $10^{-13}$,
and hence all the values in Fig.~\ref{N24M8dydz_vs_z_logarithmic}
are sufficiently far from the last ray to provide a reliable measure of
the power $p$.

\subsection{Shell collapse: Unforeseen Behavior}
\label{s3.2}

The numerical calculations also revealed a number of surprises which we
now discuss.

\begin{figure}
\includegraphics[scale=0.65]{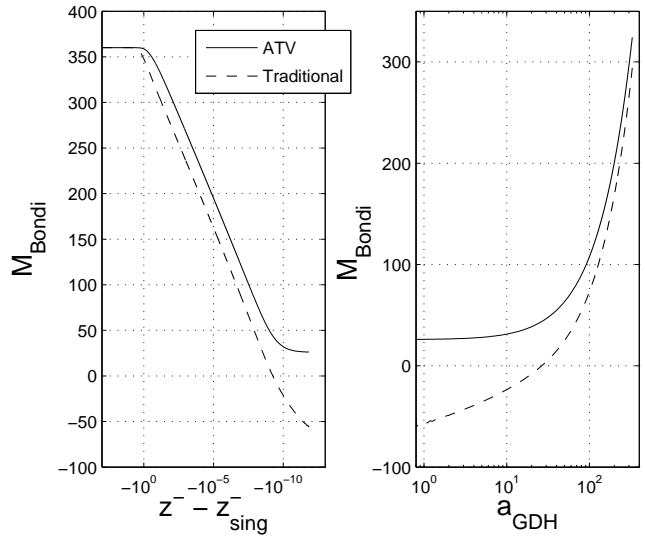}
\caption{ The ATV Bondi mass $\mbatv$ (solid lines) and the traditional
Bondi mass $\mbt$ (dashed lines) are plotted against
$z^- -z^-_{\rm sing}$ (left) and the horizon area (right). This
simulation corresponds to $M_{\ADM}=360$, $N=720$ (so $M^\star = 12$).
For high values of $N$, both formulas give a large non-zero Bondi mass
at the last ray. $\mbt$ becomes negative when the area is
still macroscopic. On the other hand $\mbatv$ remains strictly positive
all the way to the last ray, where the generalized dynamical horizon
(GDH) shrinks to zero area.}
\label{N720M360M_B_ATV_and_Trad}
\end{figure}
\emph{Bondi mass for large $\N$}: Scaling properties discussed in
section \ref{s2} imply that if the Bondi mass at the last ray is
finite, it will be macroscopic for a sufficiently large $N$. This
expectation is borne out (in particular the Bondi mass {\em is}
finite) in all our simulations with large $M_{\ADM}$ and large $\N$.
Fig. ~\ref{N720M360M_B_ATV_and_Trad} summarizes the result of a
simulation where $N =720$ and $M_{\rm ADM} = 360$ (so $\N=30$ and
$M^\star =12$). The Bondi mass, $\mbt$, that has been commonly used
in the literature {\cite{cghs,reviews,hs,sh,tu,st,dl} becomes
negative even far from the last ray when the horizon area is still
macroscopic, and has a macroscopic negative value at the last ray.%
\footnote{After this work was completed, Javad Taghizadeh Firouzjaee
pointed out to us that the fact that the traditional Bondi mass can
become negative was already noticed in \cite{tu}. Again though, in our
terminology the numerical simulation in that work corresponds to a
microscopic black hole with $M^\star = 1\, M_{\rm Pl}$.}
On the other hand, the more recent $\mbatv$ \cite{atv,vat}
remains strictly positive. As one would expect from the scaling
relations, because $N$ is large, $\mbatv$ is also macroscopic
at the last ray.
\begin{figure}
\includegraphics[scale=0.65]{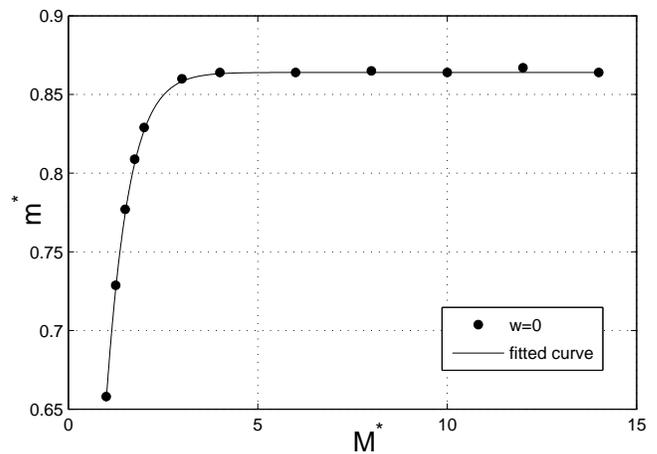}
\caption{The value of $m^\star$ (i.e. $\mbatv/\N$ at the last ray)
is plotted against $M^\star$ (which equals $M_{\ADM}/\N$) for
$M^\star \geq 1$. For Macroscopic $M^\star$ (actually, already for
$M^\star \gtrsim 4$!) $m^\star$ has a universal value, approximately
$0.864$. }
\label{mass_at_last_ray}
\end{figure}

\medskip
\emph{Universality of the end state}: Fig
\ref{mass_at_last_ray} shows a plot of $m^\star$, the value of
($\mbatv/\N$) at the last ray, against $M^\star =
(M_{\ADM}/\N)$, for several values of the initial  $M^\star
>1$. The curve that fits the data, shown in the figure, is
\be m^\star = \alpha\,(1-e^{-\beta (M^\star)^\gamma})\ee
with specific values for the constants $\alpha,\beta,\gamma$
$$ \alpha \approx 0.864,\,\, \beta \approx 1.42,\,\, \gamma\approx
1.15\,. $$
It is visually clear from the plot that there is a qualitative
difference between $M^\star \gtrsim 4$ and $M^\star \lesssim4$.
Physically this can be understood in terms of $\a_{\rm initial}$,
the area of the first marginally trapped surface: Eq (\ref{dh})
implies that $\a^\star_{\rm initial} = \a_{\rm initial}/\N$ can be
greater than a Planck unit only if $M^\star$ is larger than $3$. It
is therefore not surprising that $M^\star =4$ should serve as the
boundary between macro and Planck regimes. Indeed, as Fig
\ref{mass_at_last_ray} shows, if $M^\star \gtrsim 4$, the value of
the end point Bondi mass is universal, $m^\star \approx 0.864$. For
$M^\star \lesssim 4$ on the other hand, the value of $m^\star$
depends sensitively on $M^\star$. This could have been anticipated
because if $M^\star \le 3$, what evaporates is a \GDH\, which
\emph{starts out} with one Planck unit or less of area $\a^\star$.
\emph{Thus, in the mean-field approximation it is natural to regard
CGHS black holes with $M^\star \gtrsim 4$ as macroscopic and those
with $M^\star \lesssim4$ as microscopic.} Finally, for macroscopic
black holes, the end-value of the traditional Bondi-mass is also
universal: $\mbt < \a_{\rm hor}$ and $(\mbt/\N ) \to -2.0$ as
$\a_{\rm hor} \to 0$.

As noted in the beginning of section \ref{s3}, there have been a
number of previous numerical studies of the CGHS model
\cite{dl,ps,tu,hs}. They clarified several important dynamical
issues. However they could not unravel universality because they all
focused on cases where the black hole is microscopic already at its
inception: $M^\star \le 2.5$ in \cite{dl}, $M^\star =1$ in \cite{ps}
and \cite{tu} and $M^\star = 0.72$ in \cite{hs}. This limitation was
not noticed because the scaling symmetry and its significance was
not appreciated.

\begin{figure}
\includegraphics[scale=0.65]{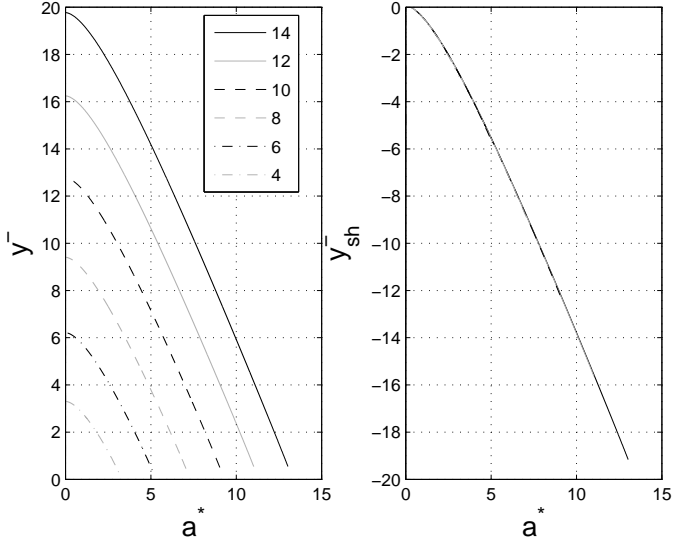}
\caption{ Left: The affine parameter $y^-$ (defined in Eq. (\ref{yz})
of the physical metric $g$
is plotted against the rescaled area $\a^\star := (\a_\GDH/\N)$ of the
generalized dynamical horizon (given by $(\Phi/\N - 2))$ at the horizon
for values of $M^\star$ from $4$ to $14$. Even though the curves are
very similar in shape, they do not coincide. Right: Once the shifting
freedom in $y^-$ is utilized, we see that a properly shifted version
$y^-_{\rm sh}$ is universal with respect to $\a^\star$ for all macroscopic
$M^\star$ values. $y^-_{\rm sh}$ can be used as a universal coordinate
similar to the horizon area.}
\label{ym_universal}
\end{figure}
\medskip
\emph{Dynamical universality of $y^-$}: The horizon area
$\a_{\GDH}$ (more precisely, its negative) provides an
invariantly defined time coordinate to test dynamical
universality of other physical quantities. Let us begin with
$y^-$, the affine parameter along $\spr$ with respect to the
physical metric $g$ defined in (\ref{yz}).
Fig.~\ref{ym_universal}, left, shows the plot of $y^-$  against
$\a^\star := (\a_{\GDH}/\N)$ for various values of $M^\star$.
These plots show that the time dependence of $y^-$ for various
values of $M^*$ is very similar but not identical. Recall,
however, that there is some freedom in the definition of the
affine parameter. In particular, in each space-time we can
shift it by a constant, and the particular value of the constant can vary
from one space-time to the next (e.g. depend on the ADM mass). This shift
does not affect any of our considerations, including the
balance law (\ref{balance}).

Let us define $y^-_{\rm sh}$ by shifting each $y^-$ so
that each solution reaches the same small non-zero value of the
horizon area, $\a^\star = \epsilon$, at the same $y^-_{\rm sh}$. It turns out that this
shift has the remarkable feature that, for initially
macroscopic black holes, all shifted curves now coincide for
\emph{all} values of $\a^\star$. Thus, we have a universal,
monotonic function of $\a^\star$ plotted in
Fig.~\ref{ym_universal}, right. Hence $y^-_{\rm sh}$ also
serves as an invariant time coordinate. In fact it has an
advantage over $\a_\GDH$: whereas $\a^\star$ is defined only
after the first marginally trapped surface is formed (see Fig
\ref{fethi-sketch}), $y^-_{\rm sh}$ is well defined throughout
the mean-field space-time $(M,g)$.

\begin{figure}
\includegraphics[scale=0.7]{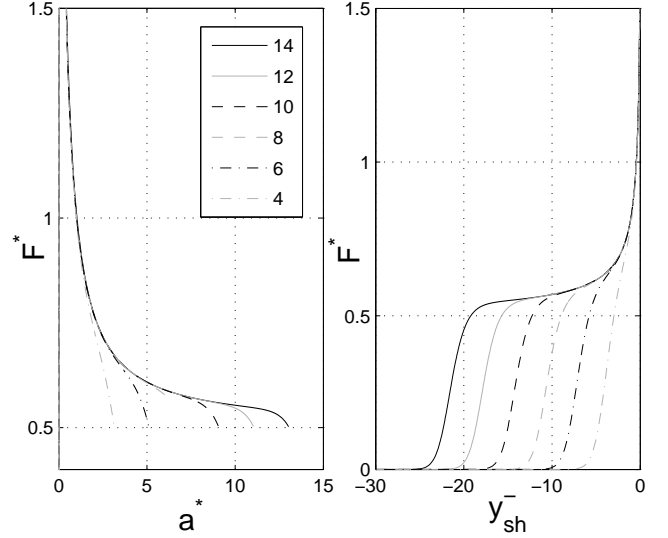}
\caption{$F^\star = (\fatv/\N)$ is plotted against the horizon area
$\a^\star:= (\a_{\GDH}/\N)$ (left) and $y^-_{\rm sh}$ (right) for
values of $M^\star$ from $4$ to $14$. For all $M^\star$ values,
$F^\star$ starts at the value of $0$ at the distant past
($\k y^-_{\rm sh} \ll -1$), and then joins a universal curve of
$F^\star$. Note that once the \GDH\, is formed, (the rightmost
beginning of each curve on the left plot) $F^\star$ is already
slightly larger in magnitude than the Hawking/thermal value 0.5
and it increases steadily as one approaches the last ray
(i.e. as $\a_{\GDH}$ and $y^-_{\rm sh}$ approach $0$.}
\label{F_ATV_universal}
\end{figure}
\begin{figure}
\includegraphics[scale=0.7]{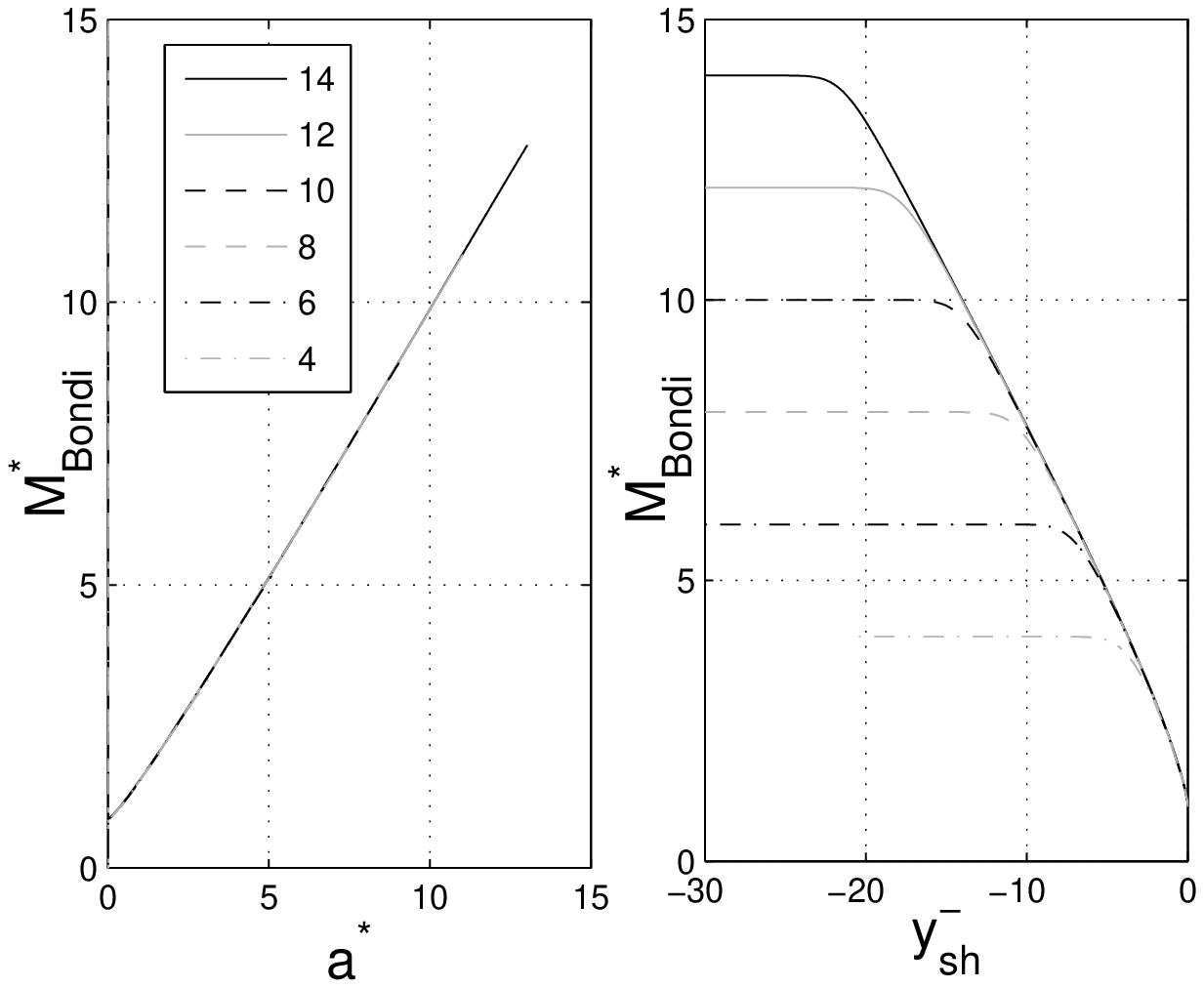}
\caption{ $M^\star_{\Bondi} = (\mbatv/\N)$ is plotted against
the horizon area $\a^\star:=(\a_{\GDH}/\N)$ (left) and $y^-_{\rm sh}$ (right)
for values of $M^\star$ from $4$ to $14$. For all these macroscopic
$M^\star$, $M^\star_{\Bondi}$ starts at the value of $M_\ADM$ in
the distant past ($\k y^-_{\rm sh} \ll -1$), and then joins a
universal curve of $M^\star_{\Bondi}$. When the dynamical
horizon first forms $M^\star_{\Bondi}$ is quite close to its initial
value of $M^\star$, (This is difficult to see in the left plot
where all the curves crowd.) This means that almost all of the
evaporation occurs after the formation of the dynamical
horizon.} \label{M_B_universal}
\end{figure}

\medskip
\emph{Dynamical Universality of $\fatv$ and $\mbatv$}: We can
repeat the procedure used above for $y^-$ to investigate if
dynamics of other physical quantities such as the Bondi flux
$F^\star := (\fatv/\N)$ and the Bondi mass $M^\star_{\Bondi}:=
(\mbatv/\N)$ are also universal. Note, however, that unlike
$y^-$, there is no `shift' (or indeed any other) freedom in
the definitions of $\fatv$ and $\mbatv$. So, if there is
universality, it should emerge directly, \emph{without any
adjustments}, in the plots of $F^\star$ and $M^\star_{\Bondi}$
against $\a^\star= (\a_{\GDH}/\N)$ or $y^-_{\rm sh}$.

Let us begin with the Bondi flux. Recall, first, that in the
external field approximation \cite{reviews,gn}, the energy flux
is very small in the distant past, rises steeply at $\k y^-
\approx - \ln(GM_{\ADM}/\k)$ and then quickly asymptotes to the
Hawking value $F^{\rm Haw} = (\N\hbar\kappa^2/2)$. This
constant flux is characteristic of thermal radiation at
temperature $\k\hbar$ in two space-time dimensions. In our
simulations (with $N=24$, or) $\bar{N}=1$ and $\hbar=\kappa=1$,
it corresponds to $F^{\rm Haw} = 0.5$.

In the mean-field theory, numerical simulations show that, for
all initially macroscopic black holes, the energy flux
$F^\star := (\fatv/\N)$ is also negligibly small in the
distant past and then rises steeply. But this rise is now
associated with a clearly identifiable dynamical process:
formation of the first marginally trapped surface. As we noted
in section \ref{s2.2}, for a shell collapse, analytical
calculations show that the area of this first surface is given
by (\ref{dh}). Assuming that we have a macroscopic initial
mass, $M^\star \gg \sqrt{G\hbar}\,\, M_{\rm Pl} =:
\tilde{M}_{\rm PL}$,\, Eq (\ref{dh}) simplifies:
\be \a_{\rm initial}^\star \,\approx\, {G\hbar}
\big[\f{M^\star}{\tilde{M}_{\rm Pl}} - 1 + \f{\tilde{M}_{\rm
Pl}}{2 M^\star} + \ldots \big]\, \ee
This relation is borne out in simulations. Assuming that the black
hole is very large at this stage, heuristically, one can equate the
area of this new born \GDH\, with the Bondi mass at the retarded
instant of time, say $y^- = y^-_o$, at which is it born. This
implies that, per scalar field, only $\sim\, 1$ Planck unit of
$M^\star_{\Bondi}$ has been radiated over the long period of time
from $y^- = -\infty$ till $y^- = y^-_o$. But once the \GDH\,
appears, the flux rises steeply to a value close to but higher than
$0.5$. Then, it joins a universal curve all the way to the last ray
where the area $\a^\star$ shrinks to zero. Thus, after a brief
transient phase around the time the \GDH\, is first formed, the
time-dependence of the Bondi flux is universal.
Fig.~\ref{F_ATV_universal}, left shows this universal time
dependence with $\a^\star$ as time and Fig.~\ref{F_ATV_universal},
right shows it with $y^-_{\rm sh}$ as time.

In virtue of the balance law (\ref{balance}) one would expect
this universality to imply a universal time dependence also for
the Bondi mass $M^\star_{\Bondi}$. This is indeed the case. At
spatial infinity $i^o_{\rm R}$,\,  we have $M^\star_{\Bondi} =
M^\star$. There is a transient phase around the birth of the
\GDH\, in which the Bondi mass decreases steeply. Quickly after
that, the time dependence of $M^\star_{\Bondi}$ follows a
universal trajectory until the last ray.
Fig.~\ref{M_B_universal}, left shows this universality with
$\a^\star$ as time while Fig.~\ref{M_B_universal}, right shows
it with $y^-_{\rm sh}$ as time.

To summarize, using either $\a^\star$ or $y^-_{\rm sh}$ as an
invariant time coordinate, we can track the dynamics of
$F^\star$ and $M^\star_{\Bondi}$. In each of the four cases,
there is a universal curve describing these dynamics. For
definiteness let us use $\a^\star$ as time and focus on
$M^\star_{\Bondi}$ (the situation is the same in the other
three cases). Since both quantities are positive, let us
consider the time-mass quadrant they span. Fix a very large
initial black hole with $M^\star= M^\star_o$ and denote by
$c_o$ the curve in the quadrant that describes its time
evolution. Then, given any other black hole with $M^\star
<M^\star_o$, the curve $c$ describing the dynamical evolution
of its $M^\star_{\Bondi}$ starts out at a smaller value of time
(i.e. $\a^\star$) marking the birth of the \GDH\, of that
space-time and, after a brief transient phase, joints the curve
$c_o$ all the way until its horizon shrinks to zero. Here we
have focused on the ATV flux and mass because their properties
make them physically more relevant. But this universality holds
also for the flux and mass expressions, $\ftrad, \mbt$ that
have been traditionally used in the literature.

\medskip
\emph{Curvature at the last ray}: There has been considerable
discussion on the nature of geometry at the last ray. Since
this ray starts out at the singularity, a natural question is
whether a curvature singularity propagates out all along the
last ray to $\spr$. This would be a `thunderbolt' representing
a singular Cauchy horizon \cite{hs}. If it were to occur, the
evolution across the last ray would not just be ambiguous; it
would be impossible. However, {\em a priori} it is not clear that a
thunderbolt would in fact occur. For, the `strength' of the
singularity goes to zero at its right end point where the last
ray originates.

Using numerical simulations, Hawking and Stewart \cite{hs}
argued that a thunderbolt does occur in the semi-classical
theory. But they went on to suggest that it could be softened
by full quantum gravity, i.e., that full quantum gravity
effects would tame it to produce possibly a very intense but
finite burst of high energy particles in the full theory.

Our calculation of the Ricci scalar very close to the last ray shows
that, except for a small region near the singularity, the scalar
curvature at the last ray is \emph{not} large
(Fig.~\ref{N24M8Ricci}). Thus, our more exhaustive and high
precision calculations rule out a thunderbolt singularity in the
original sense of the term. This overall conclusion agrees with the
later results in~\cite{dl}. (Both these calculations were done only
for initially microscopic black holes while results hold also for
macroscopic ones.) However, our calculations show that the flux
$\fatv$ does increase very steeply near the last ray (see
Fig.~\ref{F_ATV_universal}). Numerically, we could not conclude
whether the flux remains finite at the last ray or diverges.
However, the integrated flux which determines the change in
$\mbatv$ is indeed finite and in fact \emph{not very
significant} in the region very near the last ray. For
macroscopic $M^\star$ values, the total radiated energy after
the point when $F^\star$ reaches the value $1$ is $\sim 1$
Planck mass. (see Figs.~\ref{F_ATV_universal},
\ref{M_B_universal}). Thus, if we were to associate the
thunderbolt idea to the steep increase of flux at the last ray,
this would have to be in quite a weak sense; in particular,
there is no singular Cauchy horizon.

\medskip
\emph{Nature of the Bondi flux}: Recall that in the external field
approximation, the energy flux starts out very low, rapidly
increases and approaches the constant thermal value $(F^{\rm
Haw}/\N) = \hbar\k^2 /2$ ($= 0.5$ in our simulations) from below
\cite{gn,reviews}. In the mean-field theory, the flux $\fatv$ also
starts out very small and suddenly increases when the \GDH\, is
first formed. However, it overshoots the thermal value and ceases to
be constant much before the black hole shrinks to Planck size
(Fig.~\ref{F_ATV_universal}). During subsequent evolution, $\fatv$
monotonically increases in magnitude and is about 70\% greater than
the constant thermal value $F^{\rm Haw}$ when $\mbatv \sim 2\N
M_{\rm Pl}$: the standard assumption that the flux is thermal till
the black hole shrinks to Planck size is not borne out in the mean
field theory. (One's 4-dimensional intuition may lead one to think
that the increase in the flux merely reflects that the black hole
gets hotter as it evaporates; but this is not so because the
temperature of a CGHS black hole is an \emph{absolute constant},
$T_{\rm Haw} = \k\hbar$). In the interval between the formation of
the \GDH\ and the time when $\mbatv$ approaches $\N M_{\rm Pl}$, the
numerical flux is well approximated by
\be \label{fluxes} \fatv = F^{\rm Haw} \left[1 - \ln \left(1 - \f{\N M_{\rm
Pl}}{\mbatv } \right) \right] . \ee
Thus, in this interval the flux is close to the constant thermal
value only while the area $\a$ of the \GDH\ is much greater than
$\N$ Planck units.%
\footnote{The leading order correction $+\, ({\N M_{\rm Pl}}/{M_{\rm
Bondi}})$ to the Hawking flux was obtained by Ori by analytical
approximation methods and served as the point of departure for
obtaining the fit (\ref{fluxes}). Note also that if the fluxes
differ over a significant time interval, it follows that the quantum
radiation is not thermal. But the converse is not true as there are
\emph{pure} states in the outgoing Hilbert space for which the
energy flux at $\spr$ is extremely well approximated by the constant
thermal value. For quantum states, what matters is the comparison
between the function $y^-_{\rm sh}(z^-)$ and its classical
counterpart $\ub{y}^-(z^-)$ given by (\ref{uby}) \cite{atv,vat}. And
these two functions are very different. Finally, nonthermal fluxes were
also observed in a quantum model of four-dimensional spherical shell
collapse~\cite{Vachaspati:2007hr}   }
%


\subsection{Universality beyond the shell collapse.}
\label{s3.3}
\begin{figure}
\includegraphics[scale=0.65]{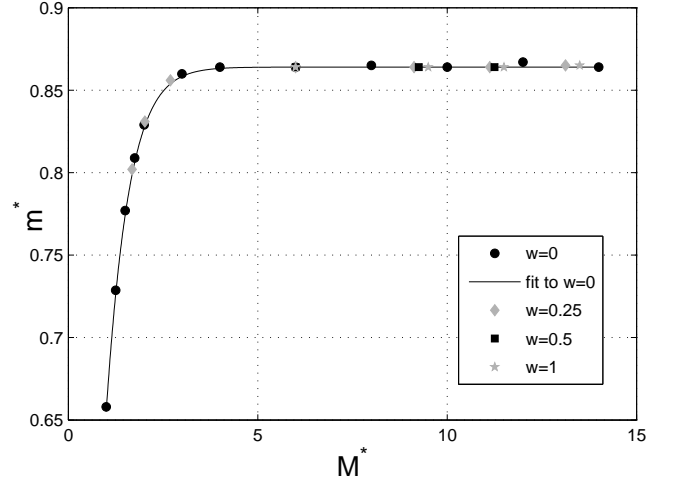}
\caption{The value of $m^\star$ (i.e. $\mbatv/\N$ at the last ray)
plotted against $M^\star$ (which equals $M_{\ADM}/\N$) for
$M^\star \geq 1$. In addition to points corresponding to shell
collapse ($w=0$) the plot now includes points with more general
profiles with $w =0.25, 0.5, 1$. The universal value
$m^\star \approx 0.864$ persists for $M^\star \ge 4$.}
\label{mass_at_last_ray2}
\end{figure}
\begin{figure}
\includegraphics[scale=0.70]{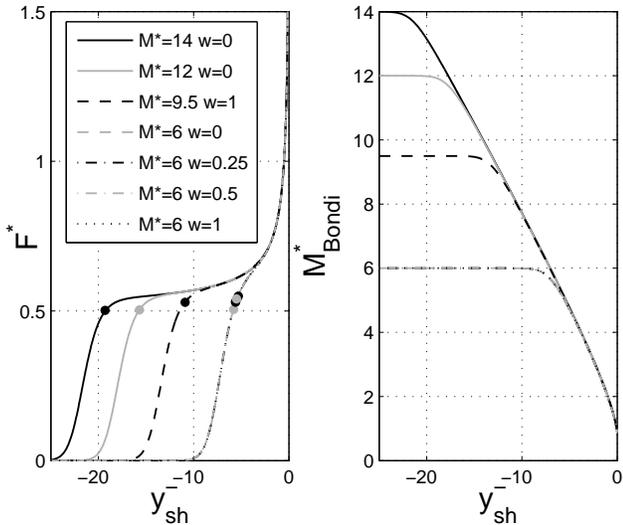}
\caption{$F^*$ (left) and $M^*_{\rm Bondi}$ (right) plotted
against $y^-_{\rm sh}$, for various incoming
matter profiles ($w$ and $M_{\ADM}$ values), including
several shell ($w=0$) cases. The time when the
dynamical horizon first forms is marked on each flux curve (which is
later for larger $w$).
All the curves with the same $M_{\ADM}$ ($6$ in this example) are on top of each other and cannot be
distinguished by the eye, showing that they have the same universal
behavior throughout the evolution, including the early times.
More generally all the asymptotic physical
quantities depend only on the combination $M_{\ADM}$ of the profile
parameters $M$ and $w$ as long as $\k w$ is small compared to the initial
area of the \GDH.}
\label{ym_universal_different_w}
\end{figure}

So far, we have focused our attention on a delta distribution shell
collapse (\ref{shell}). As we will discuss more in the following
section \ref{s4}, we expect the results to be robust for a large
class of infalling profiles, so long as the \GDH\, forms promptly.
To test this conjecture, we evolved a 2-parameter family of initial
data, parameterized by a characteristic initial mass $M$ and width
$w$. Now, it is clear from the form (\ref{sol2}), (\ref{sol3}) of
initial data that what matters is not the profile $f^{(o)}_+$ itself
but rather the integral of $(\Dp f^{(o)}_+)^2$. We will specify it
using two parameters, $M$ and $w$:
\be \label{profiles} \textstyle{\int_0^{\bar{x}^+}} \dd
\bar{\bar{x}}^+\, (\frac{\partial f_+^{(o)}}{\partial
\bar{\bar{x}}^+})^2\, = \left\{\begin{array}{cl} \f{M}{12\N}
 \left(1-e^{-\f{\left(\kappa \bar{x}^+ -1 \right)^2}{w^2}}
 \right)^4 & \bar{x}^+>1 \\
0 & \bar{x}^+<1 \end{array} \right. \ee
This choice is motivated by the following considerations. First, as
in the shell collapse, there is a neighborhood of $\sml$ in which
the solution represents the vacuum of the theory. Second, the power
$4$ on the right side is chosen to ensure high differentiability at
$x^+ =1$ (i.e. $z^+ =0$). Thus, $f_+^{(o)}$ is $\mathcal{C}^4$ and
furthermore decays faster than $e^{-C\kappa z^+}$ for any $C$ as
$z^+ \to \infty$. Third, the parameter $w$ provides a measure of the
width of the matter profile in $x^+$ coordinates, which is roughly
the width in $z^+$ for $w \, \lesssim 1$. Finally, note that we
recover the shell profile in the limit $w \rightarrow 0$ and expect
that the physical requirement of a `prompt collapse' will be met for
sufficiently small $w$. In the case of a shell profile
(\ref{shell}), the parameter $M$ represents the ADM mass. A simple
calculation shows that for a general profile in family
(\ref{profiles}), $M_\ADM$ is given by a function of the two
parameters: $M_\ADM= M(1+1.39 \ w)$. Thus, within this family, the
issue of universality of a physical quantity boils down to the
question of whether it depends only on the specific combination
$M(1+1.39 \ w)$ of the two parameters.

\begin{figure}
\includegraphics[scale=0.60]{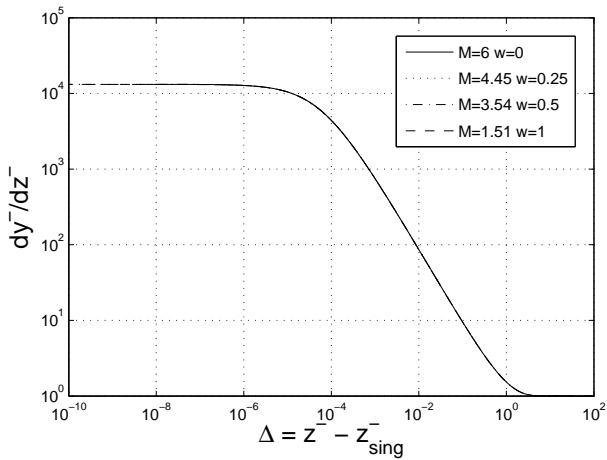}
\caption{Plot of $\dd y^-/\dd z$ against the separation
in $z^-$ from the singularity for various values of $M$ and
$w$ with a fixed ADM mass $M^\star = 6$. The functional
dependence $y^-(z^-)$ determines the physics of the outgoing
quantum state completely \cite{atv,vat}. Coincidence of
these curves in the mean-field theory suggests that the outgoing
quantum state is likely to be universal within the class of
initial data with the same ADM mass, so long as the collapse is
prompt.}
\label{dydz_vs_z_w}
\end{figure}

Numerical evolutions were carried out for $M^\star \approx
6,9,11,13$ and $w = 0.25, 0.5, 1$. We find that universality is
indeed retained for all these cases. Specifically, we repeated the
following analysis of section \ref{s3.2} for various values
of $M$ and $w$:\\
i) The relationship between the end-point values $m^\star$ of
$M^\star_{\Bondi}$ against $M^\star$; see Fig.~\ref{mass_at_last_ray2}.
For $M^\star \ge 4$, we
again find $m^\star$ has the same universal value, $\sim .864
M_{\rm Pl}$.\\
ii) The relationship of $y^-$ vs $\a^\star$ (once \GDH\
becomes time-like). As before, by an appropriate shift, we find a
$y^-_{\rm sh}$ that can be used as a
universal time coordinate for all cases.\\
iii) The dependence of $F^\star$ and $M^\star_{\Bondi}$ on
$\a^*$ and $y^-_{\rm sh}$; see Fig.~\ref{ym_universal_different_w}.
For a fixed value of $M_{\ADM}$ the plots are indistinguishable, so
that even for this broader class of matter profiles, there are two
universal curves, one for the dynamics of $F^\star$ and the other
for $M^\star_\Bondi$. In particular, for a given $w > 0$, the time
evolution $F^\star$ and $M^\star_{\Bondi}$ is identical to that
obtained with the shell collapse ($w=0$).\\

In the classical theory, if the collapsing matter $f^{(o)}_+$
is compactly supported on $\smr$, to the future of this support
the geometry is universal, determined by the ADM mass
$M_{\ADM}$. This is because stationary, classical, CGHS black
holes are characterized completely by $M_{\ADM}$. Whether the
situation would have a direct counterpart in the semi-classical
theory is not {\em a priori} clear because the semi-classical
solutions are not stationary and there is no reason to expect
the solution to be characterized just by one or two parameters
to the future of the support of $f^{(o)}$. Our results provide
a precise sense in which universality does persist. As long as
the black hole is initially macroscopic and the collapse is
prompt, we have : i) a universal asymptotic time translation
$\partial/\partial y^-_{\rm sh}$ (Fig \ref{dydz_vs_z_w}); and,
soon after the formation of the \GDH, ii) universal dynamics of
physical observables with respect to the physical asymptotic
time $y^-_{\rm sh}$.

\section{Discussion}
\label{s4}

The CGHS model provides a useful arena to explore the formation
and quantum evaporation of black holes. For, the classical
action is closely related to that governing the spherically
symmetric gravitational collapse in 4 dimensions and, at the
same time, the decoupling of matter and dilaton fields in the
model introduces significant technical simplification. However,
in this paper, we were not concerned with the \emph{full}
quantum theory of the CGHS model. Rather, we restricted
ourselves to the mean-field equations of \cite{atv,vat} and
explored their implications using high precision numerics.

\subsection{Viewpoint}
\label{s4.1}
Our analysis was carried out in the same spirit that
drove the investigation of critical phenomena in classical general
relativity \cite{mc,gmg}. There, one takes equations of general
relativity seriously and shows, for example, that black holes can
form with arbitrarily small mass. From a more complete physical
perspective, these black holes would have enormous Hawking
temperature, whence quantum effects would be crucial. To know
whether black holes with arbitrarily small masses can form in
Nature, one cannot really rely on the classical Einstein equations.
The viewpoint in those investigations was rather that, since general
relativity is a self-contained, well defined theory, it is
interesting to explore what it has to say about such conceptual
issues. The results of those explorations led to the discovery of
critical behavior in gravitational collapse, which is of great
interest from a theoretical and mathematical physics perspective. In the same vein,
in the CGHS model, it is conceivable \cite{vat} that the relative
quantum fluctuations of operators $\h\Theta,\, \h\Phi$, may become
of order $1$ once the horizon mass is of the order of, say,
$\sqrt{M^\star M_{\rm Pl}}$.
\footnote{Note incidentally that in 4-dimensions, when a black
hole with $M_{\ADM} = M_\odot$ has shrunk down through quantum
radiation to mass $\sqrt{M_{\ADM} M_{\rm Pl}}$, its horizon
radius is less than a fermi, and for a super-massive black hole
with $M_{\ADM} = 10^{9} M_\odot$, this radius is a tenth of an
angstrom.}
Suppose this were to happen at a point $p$ on the \GDH. Then, to the
future of the null ray from $p$ to $\spr$, solutions $\Theta, \Phi$
to the mean-field equations discussed in this paper would be poor
approximations of the expectation values of $\h\Theta,\h\Phi$ that
result from full quantum equations. That is, our solutions to the
mean-field equations would not be \emph{physically} reliable in this
future region. The scope of this paper did not include this issue of
the physical domain of validity of the mean-field approximation. As
in much of the literature on semi-classical gravity, we considered
the entire space-time domain where the mean-field equations have
unambiguous solutions. And as in investigations of critical
phenomena, our focus was on exploring non-trivial consequences of
these equations. Specifically, we wished to explore two questions:
\emph{Are standard expectations about predictions of semi-classical
gravity borne out? Do the mean-field dynamics exhibit any universal
features?}

\subsection{Summary and conjectures in geometric analysis}
\label{s4.2}

We found that some of the standard expectations of
semi-classical gravity are indeed borne out: The semi-classical
space-time is asymptotically flat at $\spr$ as in the classical
theory, but in contrast to the classical case $\spr$ is now
\emph{incomplete}. Thus, the expectation \cite{swh} that the
full quantum space-time would be an extension of the
semi-classical one is viable.

However, a number of other expectations underlying the standard
evaporation paradigm turned out to be incorrect. Specifically:\\
a) The traditional Bondi mass $\mbt$ is large and negative at
the end of the semi-classical evaporation rather than of Planck
size and positive;\\
b) The recently introduced Bondi mass $\mbatv$ remains positive
but is large, rather than of Planck size at the end of evaporation;\\
c) The energy flux $\fatv$ of quantum radiation deviates from
the Hawking flux corresponding to thermal radiation even when
the black hole is macroscopic, the deviation becoming larger as
the evaporation progresses; and,\\
d) Along the `last ray' from the end of the singularity to $\spr$,
curvature remains finite; there is no `thunderbolt singularity' in
the metric extending to $\spr$.

The analysis also brought out some unforeseen universalities.
The most
striking among them are:\\
i) If $M^* = M_{\ADM}/\N$ is macroscopic, at the end of
semi-classical evaporation $m^\star := \mbatv/\N$\, assumes
a universal value, $m^\star \approx .864 M_{\rm Pl}$;\\
ii) As long as $M^\star$ is greater than
$M_{\rm pl}$, there is a universal relation: $m^\star = \alpha(1-
e^{-\beta (M^\star)^\gamma})\, M_{\rm Pl}$, with $\alpha \approx
0.864$,\, $\beta \approx 1.42$,\, $\gamma\approx 1.15$;\\
iii) An appropriately defined affine parameter $y^-_{\rm sh}$
along $\spr$ is a universal function of the area $a_{\GDH}$ of
the generalized dynamical horizon;\\
iv) The evolution of the Bondi mass $\mbatv$  with respect to an
invariantly defined time parameter $\a_{\GDH}$ (or $y^-_{\rm sh}$)
follows a universal curve (and same is true for the energy flux
$\fatv$).

These results bring out a point that has not drawn the
attention it deserves: the number $N$ of fields plays an
important role in distinguishing between macroscopic and Planck
size quantities.  If semi-classical gravity is to be valid in
an interesting regime, we must have $N \gg 1$ and the ADM mass
and horizon area are macroscopic if $M_{\ADM}/\N \ge 4G\hbar
M_{\rm pl}$ and $\a/\N \ge G\hbar$. (By contrast, it has
generally been assumed that the external field approximation
should hold 
so long as $M_{\ADM} >
M_{\rm Pl}$ or $\a > G\hbar$.) Of course the ADM masses can be
much larger and for analogs of astrophysical black holes we
would have $M_{\ADM}/(\N M_{\rm pl}) \gg G\hbar$. After a brief
transient period around the time the \GDH\, is born, dynamics
of various physical quantities exhibit universal behavior till
the horizon area $\a$ goes to zero. If $M_{\ADM}/(\N M_{\rm
pl}) \gg 1$, the universal behavior spans a huge interval of
time, as measured by the physical affine parameter $y_{\rm
sh}^-$ on $\spr$ or the horizon area $\a$.

All these features are direct consequences of the dynamical
equations (\ref{de1}) and (\ref{de2}) for infalling profiles
(\ref{profiles}) characterized by two parameters $M, w$. Of
course, with numerical analysis one cannot exhaustively cover
the full range of solutions, and given the complete freedom to
specify incoming flux from $\smr$ one can always construct
initial data that will not exhibit our universal dynamics
---for example, after the \GDH\, is formed, send in a steady
stream of energy with magnitude comparable to $\fatv$. Here we have
restricted attention to initial data for which the $\GDH$ forms {\em
promptly}, and is then left to decay quantum mechanically without
further intervention. Our intuition is that universality is
associated with a \emph{pure quantum decay} of a \GDH, pure in the
sense that the decay is uncontaminated by continued infall of
classical matter carrying positive energy. Therefore, we conjecture
that for macroscopic black holes formed by smooth infalling matter
profiles of compact support, these universalities will continue to
hold soon after the \GDH\ turns time-like. More generally, for
profiles in which the positive energy flux carried across the \GDH\
by the classical fields $f^{(i)}_+$ is negligible compared to the
negative quantum flux to the future of some ray $z^+ = z^+_o$, the
universality should also hold in the future region $z^+ > z^+_o$.

This scenario provides a number of concrete and interesting
problems for the geometric analysis community. Start with
initial data (\ref{sol2}), (\ref{sol3}) at $\mathcal{I}^-$ with
$f_-^{(i)} =0$ and a smooth profile $f_+^{o}$ with compact
support for each of the $N$ fields $f^{(i)}_+$. Evolve them
using (\ref{de1}) and ({\ref{de2}). Then, we are led to
conjecture that the resulting solution
has the following properties:\\
1) The maximal solution is asymptotically flat at right future
null infinity $\spr$;\\
2) $\mathcal{I}^+_R$ is future incomplete;\\
3) A positive mass theorem holds: The Bondi mass $\mbatv$ is
non-negative everywhere on $\mathcal{I}^+_R$;\\
4) So long as $M_{\ADM} \gg \N\, \sqrt{\hbar/G}\,\kappa$, the final
Bondi mass (evaluated at the last ray) is given by $M_{\Bondi}^{\rm
final} \approx 0.864 \N\sqrt{\hbar/G}\,\kappa$;\\
5) Fix a solution $s_o$ with $M_{\ADM} = M_o \gg N_o
\sqrt{\hbar/G}\,\kappa$ and consider the curve $c_o$ describing the
time evolution of the Bondi mass in the $\a_{\GDH}/N_o$ --
$M_{\Bondi}/N_o$\, plane it defines. Then the corresponding curve
$c$ for a solution with $M/N < M_o/N_o$ coincides with $c_o$ soon
after its \GDH\, becomes time-like.

\subsection{Quantum Theory}
\label{s4.3}

Although the mean-field approximation ignores quantum fluctuations
of geometry, nonetheless our results provide some intuition on what
is likely to happen near $\spr$ in the full quantum theory. First,
because there is no thunderbolt singularity along the last ray, the
semi-classical solution admits extensions in a large neighborhood of
$\spr$ to the future of the last ray. In the mean-field
approximation the extension is ambiguous because of the presence of
a singularity along which the metric is $C^0$ but not $C^1$. But it
is plausible that these ambiguities will be resolved in the full
quantum theory and there is some evidence supporting this
expectation \cite{vat,ori}. What features would this quantum
extension have? Recall that the model has $N$ scalar fields and the
black hole emits quantum radiation in each of these channels. The
Bondi mass that is left over at the last ray is $M_{\Bondi} \approx
0.864 \N M_{\rm Pl}$. So we have $(0.864/24)M_{\rm Pl}$ units of
mass left over \emph{per channel}. It is generally assumed that this
tiny remainder will be quickly radiated away across
$\bar{\mathcal{I}}^+_{\rm R}$, the right future null infinity of the
quantum space-time that extends beyond the last ray. Suppose it is
radiated in a finite affine time. Then, there is a point $p$ on
$\bar{\mathcal{I}}^+_{\rm R}$ beyond which $\mbatv$ and $\fatv$ both
vanish. The expression (\ref{flux}) of $\fatv$ now implies that
$\bar{\mathcal{I}}^+_{\rm R}$ is `as long as' $\sml$. This is
sufficient to conclude that the vacuum state (of right moving fields
$\hat{f}^{(i)}_-$) on $\sml$ evolves to a pure state on
$\bar{\mathcal{I}}^+_{\rm R}$ (because there are no modes to trace
over). This is precisely the paradigm proposed in \cite{atv}. Thus,
the semi-classical results obtained in this paper provide concrete
support for that paradigm and re-enforces the analogous
4-dimensional paradigm of \cite{ab} (which was later shown to be
borne out also in the asymptotically AdS context in string theory
\cite{hls}).

All our analysis was restricted to the 2-dimensional, CGHS black
holes. As we mentioned in section \ref{s1}, while they mimic several
features of 4-dimensional black holes formed by a spherical
symmetric collapse of scalar fields, there are also some key
differences. We will conclude with a list of the most important of
these differences and briefly discuss their consequences. (For a
more detailed discussion, see \cite{vat}.)

First, in 2-dimensions, surface gravity $\k$ and hence, in the
external field approximation, the Hawking temperature $T_{\rm Haw}$,
is a constant of the theory; it does not depend on the specific
black hole under consideration. In 4-dimensions, by contrast, $\k$
and $T_{\rm Haw}$ depend on the black hole. In the spherical case,
they go inversely as the mass so one is led to conclude that the
black hole gets hotter as it evaporates. A second important
difference is that, in the CGHS black hole, matter fields $f^{(i)}$
are decoupled from the dilaton and their propagation is therefore
decoupled from the dynamics of the geometric sector. This then
implies that the right and left moving modes do not talk to one
another. In 4 dimensions, the $f^{(i)}$ are directly coupled to the
dilaton and their dynamics cannot be neatly separated from those of
geometric fields $\Phi, \Theta$. Hence technically the problem is
much more difficult. Finally, in 4 dimensions there is only one
$\mathcal{I}^+$ and only one $\mathcal{I}^-$ while in 2 dimensions
each of them has two distinct components, right and left.
Conceptually, this difference is extremely important. In 2
dimensions the infalling matter is only in the plus modes,
$f^{(i)}_+$, and its initial state is specified just on $\smr$ while
the outgoing quantum radiation refers to the minus modes,
$f^{(i)}_-$, and its final state has support only on $\spr$. In
4-dimension, there is no such clean separation.

What are the implications of these differences?

Because of the first two differences, analysis of CGHS black holes
is technically simpler and this simplicity brings out some features
of the evaporation process that can be masked by technical
complications in 4 dimensions. For instance, since the Hawking
temperature $T_{\rm Haw}$ is an absolute constant ($\hbar\k$) for
CGHS black holes, the standard paradigm that the quantum radiation
is thermal till the black hole has shrunk to Planck size leads to a
clean prediction that the energy flux should be constant, $F^{\rm
Haw} = \hbar\k^2/48$. We tested this simple prediction in the mean
field approximation and found that it does not hold even when the
horizon area is macroscopic. In 4 dimensions, since the temperature
varies as the black hole evaporates, testing the standard paradigm
is much more delicate. Similarly, thanks to the underlying technical
simplicity in the CGHS case, we were able to discover scaling
properties and universalities. We believe that some of them, such as
the `end point universality', will have counterparts in 4 dimensions
but they will be harder to unravel. The CGHS results provide hints
to uncover them.

The third difference has deeper conceptual implications which we
will now discuss in some detail. In 4-dimensions, since there is a
single $\mathcal{I}^-$ and a single $\mathcal{I}^+$, unitarity of
the quantum S-matrix immediately implies that all the information in
the incoming state can be recovered in the outgoing state. In 2
dimensions, on the other hand, there are two distinct questions: i)
is the S-matrix from $\sml$ to $\spr$ unitary? and ii) is the
information about the infalling matter on $\smr$ recovered in the
outgoing state at $\spr$? As discussed above, results of this paper
strongly support the paradigm of \cite{atv,vat} in which the answer
to the first question is in the affirmative; information on $\sml$
is faithfully recovered on $\spr$. However, this does \emph{not}
imply that all the infalling information at $\smr$ is imprinted on
the outgoing state at $\spr$.

In the early CGHS literature, this second issue was often mixed with
the first one. Because it was assumed that all (or at least most) of
the ADM mass is evaporated away through quantum radiation, it seemed
natural to consider seriously the possibility that all the
information in the infalling matter at $\smr$ can be recovered from
the outgoing quantum state at $\spr$. The key question was then to
find mechanisms that make it possible to transfer the information in
the \emph{right-moving} infalling modes $f^{(i)}_+$ to the
\emph{left-moving} modes $f^{(i)}_-$ going out to $\spr$. In
\cite{st}, for example, the 2 dimensional Schwinger model with a
position dependent coupling constant was discussed in some detail to
suggest a possible mechanism.

However, our universality results strongly suggest that these
attempts were misdirected. The physical content of the outgoing
\emph{quantum} state is encoded entirely in the function $y^-_{\rm
sh}(z^-)$ \cite{atv,vat} on $\sprb$, the right future null infinity
of the quantum extension of the semi-classical space-time. In the
family of profile functions $f^{(o)}_+$ we analyzed in detail, the
function $y^-_{\rm sh}(z^-)$ on $\spr$ has universal behavior,
determined just by the total ADM mass. Since only a tiny fraction of
Planck mass is radiated per channel in the portion of $\sprb$ that
is not already in $\spr$, it seems highly unlikely that the
remaining information can be encoded in the functional form of
$y^-_{\rm sh}(z^-)$ in that portion. Thus, at least for large
$M^\star$ we expect the answer to question ii) to be in the
negative: information contained in the general infalling matter
profile on $\smr$ will \emph{not} be fully recovered at $\spr$. From
our perspective, this is not surprising because the structure of
null infinity in the CGHS model is rather peculiar from the
standpoint of 4 dimensions where much of our intuition is rooted. In
2 dimensional models, $\sprb$ is not the \emph{full} future boundary
of space-time. Yet discussions of CGHS black holes generally ignore
$\szpl$ because, as we saw in section \ref{s1}, even in the
classical theory the black hole interpretation holds only with
reference to $\spr$. Indeed, for this reason, investigations of
quantum CGHS black holes have generally focused on the Hawking
effect and question i) of unitarity, both of which involve dynamics
only of $\hat{f}_-^{(i)}$ for which $\sprb$ does effectively serve
as the complete future boundary.

In 4 dimensions, the situation is qualitatively different in this
regard: in particular, the outgoing state is specified on all of
future null infinity $\mathcal{I}^+$, not just on half of it.
Therefore, if the singularity is resolved in the full quantum
theory, $\sprb$ \emph{would} be the complete future boundary of the
quantum space-time and there would be no obstruction for the $S$
matrix to be unitary and hence for the full information on
$\mathcal{I}^-$ to be imprinted in the outgoing state on
$\mathcal{I}^+$.

\bigskip\bigskip

\textbf{Acknowledgements:}\, We would like to thank Amos Ori,
Madhavan Varadarajan for discussions and for their comments on our
manuscript. This work was supported by the NSF grants PHY-0745779,
PHY-0854743, the Eberly research funds of Penn State, and the Alfred
P. Sloan Foundation.

\end{document}